\documentclass[5p,times]{elsarticle}

\usepackage[utf8]{inputenc}
\usepackage{verbatim}
\usepackage{natbib}
\usepackage{algorithmicx}
\usepackage{svg}
\usepackage{xcolor}
\usepackage{caption}

\usepackage{epstopdf}
\usepackage{booktabs}
\usepackage{enumitem}
\usepackage{amsmath,amssymb,amsfonts}
\usepackage{graphicx}
\usepackage{textcomp}
\usepackage{appendix}
\usepackage{latexsym}
\usepackage{algorithm}
\usepackage{booktabs}
\usepackage{tabularx}
\usepackage{courier}
\usepackage{listings}
\usepackage{xcolor}
\usepackage{color,soul}
\usepackage{amsthm}
\usepackage{multirow}
\usepackage{comment}
\usepackage{todonotes}
\usepackage{pifont}% http://ctan.org/pkg/pifont
\usepackage{algpseudocode}% http://ctan.org/pkg/algorithmicx
\algrenewcommand\textproc{}% Used to be \textsc
\usepackage[all]{nowidow}
\usepackage{array}
\usepackage{pifont}% http://ctan.org/pkg/pifont
\usepackage[skip=0pt,font=small]{caption}
\usepackage{bm}
\usepackage{tikz}
\usepackage{soul}
\usepackage{xcolor}
\usepackage{algorithm} 
\usepackage{algpseudocode} 
\usepackage{tabularx}
\usepackage{breqn}
\usepackage{mathtools} 
\usepackage{makecell}
\usepackage{url}
\usepackage[switch]{lineno}
\nolinenumbers
\newcommand{\algo}[0]{\mbox{\textsc{ParDNN}}}

\newcommand{\wahib}[1] {{\color{green}{Wahib}}: {\color{red}{#1}}}
\newcommand{\edit}[1] {{\color{green}{}} {\color{black}{#1}}}

\begin{document}
\sloppy
\title{A Computational-Graph Partitioning Method for Training Memory-Constrained DNNs\\
}

\author[address]{Fareed Qararyah}
\author[address2]{Mohamed Wahib}
\author[address3]{Doğa Dikbayır}
\author[address4]{Mehmet Esat Belviranli}
\author[address]{Didem Unat}

\address[address]{Ko\c{c} University, Turkey}
\address[address2]{National Institute of Advanced Industrial Science and Technology, Japan}
\address[address3]{Michigan State University, USA}
\address[address4]{Colorado School of Mines, USA}

\begin{abstract}
Many state-of-the-art Deep Neural Networks (DNNs) have substantial memory requirements. Limited device memory becomes a bottleneck when training those models. We propose \algo{}, an automatic, generic, and non-intrusive partitioning strategy for DNNs that are represented as computational graphs. \algo{} decides a placement of DNN’s underlying computational graph operations across multiple devices so that the devices' memory constraints are met and the training time is minimized. \algo{} is completely independent of the deep learning aspects of a DNN. It requires no modification neither at the model nor at the systems level implementation of its operation kernels. \algo{} partitions DNNs having billions of parameters and hundreds of thousands of operations in seconds to few minutes. Our experiments with TensorFlow on $16$ GPUs demonstrate efficient training of $5$ very large models while achieving superlinear scaling for both the batch size and training throughput. \algo{} either outperforms or qualitatively improves upon the related work.

\end{abstract}

%% 2012 ACM Computing Classification System (CSS) concepts
%% Generate at 'http://dl.acm.org/ccs/ccs.cfm'.

%% End of generated code

%% Keywords
%% comma separated list

%% \maketitle
%% Note: \maketitle command must come after title commands, author
%% commands, abstract environment, Computing Classification System
%% environment and commands, and keywords command.
\maketitle

\footnote{This manuscript version is made available under the CC-BY-NC-ND 4.0 license http://creativecommons.org/licenses/by-nc-nd/4.0/}
\section{Introduction}
\label{sec:intro}
%opening
Deep Learning (DL) is being increasingly applied in a wide range of scientific and engineering domains. DNNs have doubled in size roughly every ${2.4}$ years due to the ability of larger models, {\em deeper} or {\em wider} or both, to produce results with higher accuracy on more complex tasks~\cite{goodfellow2016deep}. This growth is expected to continue in the coming years~\cite{Rajbhandari2019ZeROMO,4c7ab9026888dfa485b3486cfab857fc}. The deepening and/or widening of these models comes at a cost of larger memory required to store the parameters and the intermediate results\cite{sekiyama2018profile}. 
An example from computer vision field is Wide Residual Network~\cite{zagoruyko2016wide}, a widened variant of the well-known Resnet~\cite{he2016deep}, widening the model $8$ times increases the number of its parameters $\sim60$ times~\cite{zagoruyko2016wide} leading to a substantial increase in the memory requirements. The same trend shows up in the NLP field where deep-stacked LSTMs~\cite{wu2016google} or attention layers~\cite{NIPS2017_7181} often give more accurate results compared to shallower models. Introducing residual connections among the layers in a stack enabled the training of very deep encoder and decoder networks, e.g. larger versions of the Transformer model~\cite{NIPS2017_7181}, with newer models pushing the number of parameters up to $O(10B)$~\cite{shoeybi2019megatron,Rajbhandari2019ZeROMO}.

Different approaches have been proposed to tackle the issue of training very large models on multiple devices.
One approach is to work on the model level, where the model is partitioned across multiple devices through model, pipeline, channel parallelism, or combinations of them ~\cite{gholami2017integrated,jia2018beyond,Naoya:sc19,huang2018gpipe,narayanan2019pipedream, shoeybi2019megatron, jia2018exploring}. Even though these methods are successful to some extent, they suffer from either: (a) being not generic as they target a specific class of DNNs,
%: while ~\cite{shoeybi2019megatron} is targeting structure of transformer networks,~\cite{Naoya:sc19,jia2018exploring} are dealing with convolutional DNNs, 
(b) introduce non-negligible memory overhead %~\cite{narayanan2019pipedream} 
to maintain the statistical efficiency, or (c) can incur a high implementation cost and necessitate detailed  understanding of the DNN model for an accurate cost model. 
%to drive the partitioning ahead of time. %~\cite{huang2019gpipe}.
Another approach works at the systems level by partitioning the computational graph that represents the operations in a neural network model and distributes it over multiple devices. The existing work in this direction has some limitations. The method proposed in \cite{wang2019supporting} has a restricted applicability because it relies on a descriptive language to specify computations and cannot describe all the operations used in DL. Others propose a reinforcement learning-based approach, which is impractical in many cases due to substantial resource and time requirements~\cite{mirhoseini2018hierarchical,10.5555/3305890.3305932}.~\edit{In \cite{mayer2017tensorflow} authors propose a set of practical, generic, and low overhead heuristics to partition the DNN graph. They concluded that critical path-based approaches yield the best performance. However, their evaluation is based on an event-based simulation rather than on an actual DL framework.} 

We adopt the system-level approach and propose  a generic, efficient, and non-intrusive partitioning strategy (\algo) that avoids the drawbacks of the related work. \algo{} directly works on the computational graph %, a system-level
representation of the neural network adopted by the most popular general-purpose DL frameworks such as TensorFlow~\cite{abadi2016tensorflow} and MXNet \cite{chen2015mxnet}. Operating on the graph level has three main benefits. First, it provides a fine-grained view of the model, which gives more parallelization options and allows better load balancing and resource utilization. Second, it isolates our strategy from the details of the learning process, what provides more generality and guarantees unaffected statistical efficiency~\cite{narayanan2019pipedream} of the model. 
%is not going to be affected. 
Third, working at the level of the graph enables us to leverage decades of work on graph partitioning and static scheduling (as will be discussed later). %For instance, in this work we build upon a large body of work in cluster mapping and linear clustering~\cite{wang2018list}.

%derive an optimized method for balanced partitioning of DNN training DAGs. 
%Unlike prior work~\cite{wang2019supporting}, 
%\algo{} does not require any changes to the implementation of the operation-kernels in the underlying graph. 

\begin{comment}
Compile-time allocation of a set of operations represented as a directed graph, depicting the logical and temporal dependencies among them, to a set of processing elements aiming at reducing the execution time is intensively studied under the title of static task graph scheduling. Plenty of sophisticated and high-quality algorithms were proposed~\cite{kwok1995bubble, kwok1996dynamic, yang1994dsc, hwang1989scheduling, he2018novel}. The vast majority of these algorithms were developed in 90's to handle small-sized graphs, and they  were later evaluated using instances having tens or hundreds of nodes; less frequently exceeding 1000 and at maximum 3000 nodes~\cite{wang2016comparative, liou1997comparison, he2018novel,wang2018list, gerasoulis1992comparison}. A recent evaluation on large graphs shows that they either do not scale due to their high time-complexity, or produce lower quality allocations due to their local nature and inability to capture the global structure of the graph~\cite{ozkaya2019scalable}. To overcome these two limitations 
\end{comment}

%what KCPP does
\algo{}'s strategy is composed of two main steps. First, we cluster the operation-nodes of the computational graph into $K$ partitions, where $K$  represents the number of the available devices. The objective of this step is to reduce the end-to-end runtime by assigning the operations on the partitions such that the computational loads are balanced and the communication is minimized.
%\didem{next sentences are related work, here discuss our approach not related. This is not the right place. I am commenting them out and moving them to related works}
In the second step, we check whether the memory constraints are met in each partition. If they are not, we reassign some operations to different partitions such that the reassigned operations have the least possible perturbed effect on the placement generated by the first step but at the same time meet the memory constraints.

%To partition graphs having hundreds of thousands of nodes in a reasonable time, \algo{} is designed to use a set of simple and efficient heuristics instead of a single sophisticated yet time-consuming algorithm. 
%\didem{should we remove the following paragraph?}

Most existing graph partitioning libraries are designed to handle undirected graphs. Extensive experimenting done in ~\cite{10.5555/3305890.3305932, mirhoseini2018hierarchical} with state-of-the-art graph partitioning-based tools, such as Scotch static mapper ~\cite{pellegrini1996scotch, pellegrini2009distillating} and MinCut optimizer, shows that they result in $2$ to $10$ times slowdown when applied on directed graphs of DL models.
Our algorithm outline is inspired by the principle of the multilevel approach used in graph partitioning~\cite{karypis1995multilevel} but the design and algorithmic details of \algo{} includes a mix of variants of static scheduling heuristics~\cite{kim1988general} that are mutated to reduce the time complexity, and novel techniques to address some shortcomings in the existing ones~\cite{mccreary1996problem,radulescu1998glb}. 
Our contributions are:
\begin{itemize}
  \item We propose a novel computational graph partitioning method that enables training models with large memory consumption on a set of devices with limited memory.
  %\item \didem{will edit, why mention 4 gpus? }
  %\item \ppopp{We conduct extensive experiments with large DNNs from the areas of computer vision, video prediction, language modeling, and translation to demonstrate \algo{}'s efficiency. In comparison to related work that overcomes the memory capacity, \algo{} outperforms model parallelism (Mesh-TensorFlow~\cite{Mesh-TF}) and qualitatively improves upon redundant recomputation approaches~\cite{10.1145/3373376.3378505}}.    
    \item \edit{We conduct extensive related work comparisons with large DNNs:
    %computer vision, video prediction, language modeling and translation 
    %to demonstrate \algo{}'s efficiency. 
    %In comparison to related work: 
    (a) \algo{} outperforms other graph-based approaches such as linear clustering~\cite{kim1988general} and a critical path-based method~\cite{mayer2017tensorflow}, (b) \algo{} outperforms Mesh-TensorFlow, a state-of-the-art distributed training framework~\cite{Mesh-TF} as well as having qualitative advantages over it by automating the partitioning and not requiring model rewrite. (c) It generally outperforms redundant recomputation methods (Gradient Checkpointing~\cite{10.1145/3373376.3378505}). 
    %(d) It outperforms out-of-core methods (CUDA Unified Memory).
    }
%    qualitatively improves upon redundant recomputation approaches~\cite{10.1145/3373376.3378505}}.  
  
  \item For models that do not fit into a single GPU's memory, \algo{} enables training models having up to $5.1$ billion parameters using only $4$ GPUs. For models that barely fit into a single device memory, it allows more efficient training by superlinearly scaling the batch size, and in many cases, the training throughput.
  %While scaling up the batch size we constantly achieved speedups, which are super-linear in many cases as well.  %Fareed, not used.Moreover, we experimented with small models while limiting the available device memory to diversify our test set and demonstrate the generality of our approach.%
  %\didem{it will confuse the reader when we say we modified TF runtime as we claimed earlier that we don't make any changes to TF.}
  \item \algo{}'s overhead is negligible. For a graph having hundreds of thousands of nodes representing DNNs with billions of parameters, it takes $\sim 2$ minutes to find a partition for 16 GPUs, while training these models takes days or even weeks.

\item To the best of our knowledge \algo{} is the first of its type that permits the training of models that do not fit into a single device memory while being generic due to (a) having zero dependency and requiring no knowledge about the DL aspects of the models, and (b) not requiring any modifications of the model or the operation kernels.
%or usage of specific implementations .

%\item \algo{} has the potential to serve as static scheduler for large task graphs. Theoretical comparison between \algo{} and the famous Earliest Times First (ETF)~\cite{hwang1989scheduling} static scheduling algorithm, demonstrate \algo{} .....  
  %\item We propose level aware load balancing (LALB), a novel cluster mapping heuristic that can be used with static scheduling algorithms.\fareed{...needs discussion}
\end{itemize}
\section{Background}
\label{sec:Background}

Modeling a computation as a directed graph has been adopted in scheduling theory~\cite{sinnen2007task}, in parallel programming and run-time environments~\cite{lam1991coarse,yang1992pyrros,newton1992code,augonnet2011starpu}, and recently in DL frameworks \cite{abadi2016tensorflow,chen2015mxnet,bergstra2011theano}. 
TensorFlow uses a stateful dataflow graph to represent a computation. It extends the classical dataflow graph model to allow maintaining and updating the persistent state of some special nodes, branching, and loop control. In a TensorFlow graph ${G = (V, E)}$, each node ${n \in V}$ represents the instantiation of an operation (e.g., matrix multiplication or convolution) and it has zero or more inputs and zero or more outputs. Each edge ${e \in E}$ represents a dependency between its incident nodes. Normal edges represent the data flowing between the nodes, while special edges, e.g. control dependencies, are used to enforce happens-before relationships with no data flows along them~\cite{abadi2016tensorflow}.

%\subsection{Task Graph Scheduling and Graph partitioning}
%Allocating a task graph nodes on a finite set of processors to minimize the end-to-end execution time is known as task graph scheduling problem. 

Graph partitioning is, in general, defined as splitting the graph ${G(V, E)}$ into ${K}$ disjoint subsets~\cite{bichot2011graph}.
%${\{P_1,..., P_K\}}$ such that  $\cup^{K}_{k=1}P_k = V$, ${ \forall i \{1,...,K\}: P_i \neq \phi}$ and ${\forall(i,j) \in \{1,...,K\}^2, i \neq j, P_i \cap P_j = \phi } $~\cite{bichot2011graph}. 
The constrained version of the graph partitioning aims at partitioning in such a way that the sums of the vertices weights in each set are as equal as possible, and the sum of the weights of edges crossing between sets is minimized~\cite{karypis1995multilevel}. An extension of general graph partitioning which aims to assign a set of communicating tasks to processors is called {\em static mapping}~\cite{bichot2011graph}. Static mapping does not consider the logical and temporal dependencies of the tasks, it is assumed that all the tasks simultaneously coexist throughout the program execution.

%A schedule ${S}$ of a dataflow graph ${G = (V, E)}$ on a finite set ${PE}$ of processing elements is the function pair ${(ts, proc)}$, where  ${ts: V \rightarrow Q_0^+}$ is the start time function of the nodes of ${G}$; and ${proc: V \rightarrow PE}$ is the processor allocation function of the nodes of ${G}$ to the $K$ processing elements. The scheduling problem is to determine a feasible schedule ${S}$ of minimal length for ${G}$ on ${PE}$ ~\cite{sinnen2007task}. Where the schedule length, {\em makespan}, is the completion time (${C_t}$) of the last node in ${G}$  assuming that the execution of the graph starts at time ${0}$. The goal is to minimize ${C_{t max}}$, where ${C_{t max} = max_{n \in V} C_t(n)}$. 

%\didem{Converted $P$  to PE and k to K to be consistent with the table.}
%\didem{related work deserves to have its own section. }

Finding a spatial and temporal assignment of the set of nodes in a task graph $G=(V, E)$ onto a set of processors resulting in the fastest possible execution, while respecting the precedence constraints expressed by all ${e \in E}$ is referred to as task scheduling problem~\cite{sinnen2007task}. The schedule length, {\em makespan}, is the completion time (${C_t}$) of the last node in ${G}$  assuming that the graph execution starts at time ${0}$.~\edit{Where $C_t(n)$ is the time required to execute the operation represented by $n$ added to the time at which this operation starts to execute.}~The goal is to minimize ${C_{t max}}$, where ${C_{t max} = max_{n \in V} C_t(n)}$. Finding an optimal schedule or static mapping is {\em NP-hard}~\cite{bichot2011graph,sinnen2007task}. 

\section{\algo{}: A Partitioning Strategy for DNNs}
\label{sec:pardnn}
\algo{} works at the computational graph level and offers a practical, non-intrusive, and generic method to partition a neural network model on a set of processing elements(${PE}$).

 The main objective of~\algo{} is to minimize ${C_{t max}}$, the makespan of the graph, while satisfying the memory capacity constraints of the target processing elements. It is important to mention that \algo{} does not have a runtime component. All the steps of \algo{} are done ahead of time. After running \algo{} once, the resulting partitioning can be used as long as the model parameters that affect the memory consumption do not change.
 
  \begin{figure}
 \begin{center}
 \includegraphics[scale=0.45]{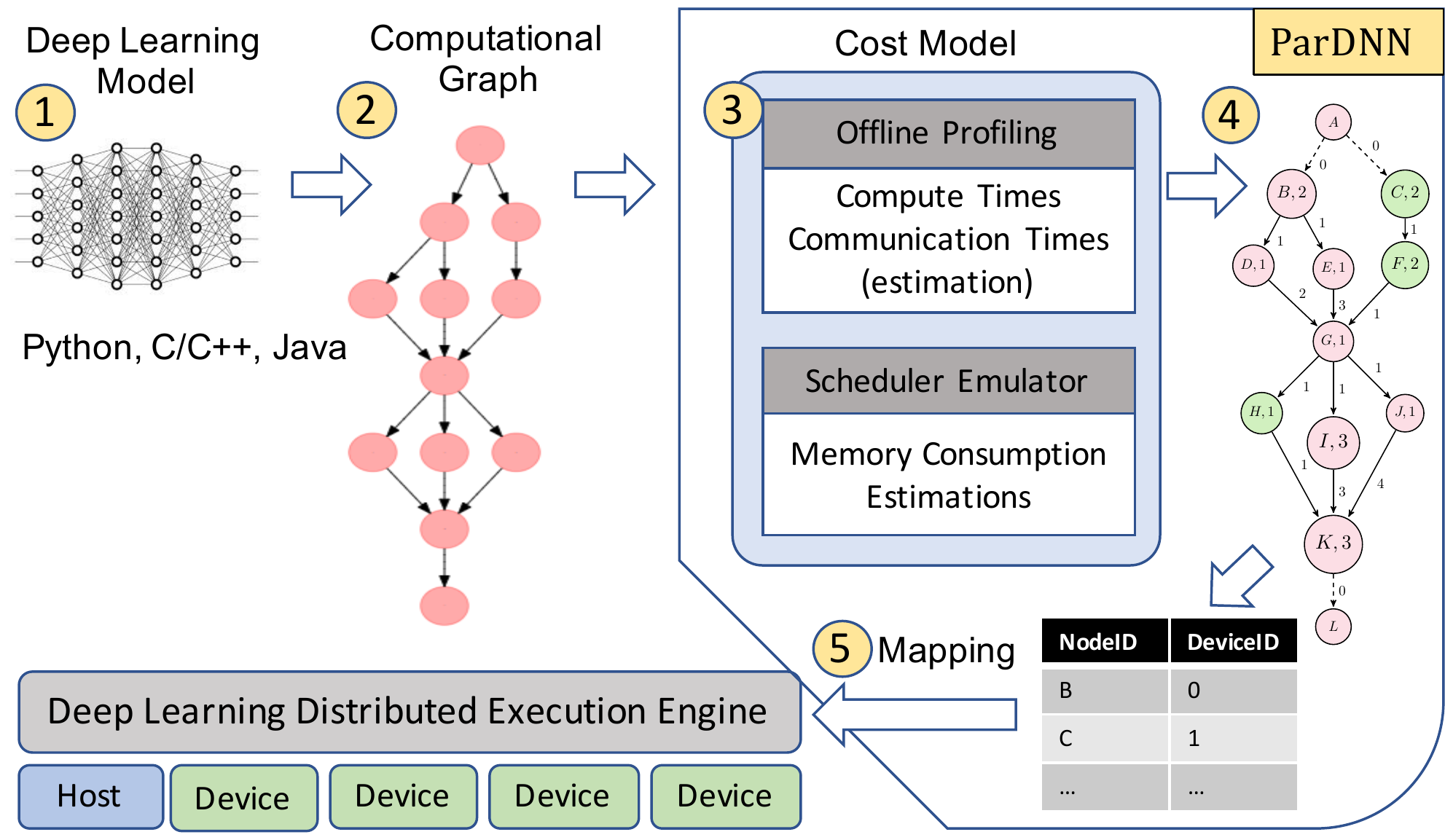}

  \caption{\algo{} Overview}
 \label{fig:imp}
 \end{center}
 \end{figure}
 
  \begin{table}[t]
\scriptsize{
\caption{Complexity of Each Step of \algo{}}
%\wahib{I am assuming we add up the steps since they are sequential, I can understand that most will add up to the overall, except the mapping. Is there a proof that $K|V|+|E|$ is always lger than the mapping complex? and even then, shouldn't we use complexity bounds? or do you have something else in mind?} \wahib{Update: I changed it to ${O(\vert V \vert * log\vert V \vert) + K(\vert E \vert)}$, until we discuss again and decide otherwise} 
\label{tab:complexity}
\begin{tabularx}{\linewidth}{|l|X|}
\hline
Step-1 & Partition to Minimize Makespan \\\hline
\quad Graph Slicing (inc. sorting) & ${O(K(\vert V \vert + \vert E \vert))}$ \\
\quad Mapping & ${O(\vert V \vert * log^2\vert V \vert)}$ 
\\\hline
%\quad Refinement & ${O(\vert V \vert*log(\vert V \vert) + K \vert E \vert)}$ 
Step-2  & Memory Heuristic - I   \\\hline
\quad TensorFlow Scheduler Emulator & ${O(\vert V \vert+ \vert E \vert)}$ \\
\quad Memory Consumption Tracker & ${O(\vert V \vert)}$ \\
\quad Overflow Handler & ${O(\vert V^2 \vert)}$ \\\hline
\edit{Step-2 } & Memory Heuristic - II \\\hline
\quad \edit{Residual Nodes Movement and CP splitting} & ${O(\vert V \vert)}$ \\\hline
\edit{Overall \algo{} Complexity (w. Heuristic-I)}& ${O(\vert V \vert^2)}$\\\hline
\edit{Overall \algo{} Complexity (w. Heuristic-II)} & ${O(\vert V \vert*log^2(\vert V \vert) + K \vert E \vert)}$\\\hline

\end{tabularx}
}
\end{table}

Figure \ref{fig:imp} shows the overall process. \algo{} takes a computational  directed acyclic graph as an input, it annotates this graph with computation, communication, and memory consumption information gathered using offline profiling. It adds an artificial source and sink node to the graph. Using the collected data,  \algo{} splits the graph into parts to be mapped to processing elements. \algo{} outputs the mapping information to be used by the execution engine of the DL framework (e.g., TensorFlow).

Our strategy is divided into two major steps. Step-1 aims to obtain a partitioning that has a minimal makespan.
Step-1 is further divided into two stages. Stage-I, {\em graph slicing} splits the graph into $K$ disjoint primary and $S$ disjoint secondary clusters. 
%A secondary cluster is a linear cluster $lc$ (Table \ref{tab:terms}). 
This splitting enables working at a coarser level in the upcoming stages. Stage-II, {\em mapping}, merges these $S$ secondary clusters into the $K$ primary clusters using a novel mapping algorithm.~\edit{In Step-2, we propose two alternative heuristics to overcome the memory overflow; one is threshold-based and the other is balancing-based. In heuristic-I, the result from Step-1 is validated against the memory constraints of the given devices. If the memory constraints are satisfied, 
%condition in Eq. \eqref{eq:1} is satisfied, 
the partition will be the final output. Otherwise, nodes are moved between the partitions until the memory consumption by a processing element $pe$ at any time $t \in [0, C_{t max}]$ is less than or equal to $pe$'s memory capacity. On the other hand, heuristic-II tries to balance the memory consumption across the available processing elements. 
Heuristic-I is fully applied in Step-2 to avoid higher time and code complexity; applying it in Step-1 requires continuous backtracking as the movement of nodes affects the temporal load balance and communication cost. Moreover, this gives a global view of the graph and wider range of moves to mitigate the memory overflow. 
Heuristic-II is partially done in Step-1 as explained in Section~\ref{sec:heu2}.
}

%\begin{equation}
%\label{eq:1}
%Max_{(t \in [0, C_{t max}],\ pe \in PE)} M_{cons} (pe, t) \leq M_{capacity}(pe) 
%\end{equation}

Table~\ref{tab:complexity} summarizes the time complexity of each step of \algo{}. The reported complexities after each step are relaxed ones and for some stages a tighter bound maybe driven with amortized analysis. 
%Splitting the partitioning strategy into a set of simple, yet efficient, sub-stages permits lowering the complexity. The nodes are grouped into clusters in the first step, then for the most of later stages, \algo{} works at the cluster rather than node granularity, which considerably reduces the instance size it deals with. 
In practice, the average running time of \algo{} on the DNN models listed in Table~\ref{tab:models} is roughly 2 minutes on a typical laptop processor, namely an Intel i7-7600u CPU @ 2.80GHz. Considering the training time of those models is in the orders of days or even weeks, \algo{} offers an extremely lightweight and practical approach to partition the DNN graphs.

Next, we explain the details of each step along with the time complexity. Table~\ref{tab:terms} summarizes the terms and notations for the explanations. %Table ~\ref{tab:complexity} shows the complexity of each algorithmic step, which will be detailed in the subsequent sections. 

\begin{table}
\caption{Terminology used in this work}
\footnotesize{
\label{tab:terms}
\begin{tabularx}{\linewidth}{c|X}
Term & Description \\
\hline
$G=(V,E)$ & Computational graph with vertex set $V,$ edge set $E$  \\\hline
$CP$ & Critical path of a graph, the longest path in the graph considering computation and communication costs. \\\hline
%$src$ & Source node of a graph\\\hline
%$snk$ & Sink node of a graph\\\hline
$C_{t max}$ & Makespan of $G$, schedule length\\\hline
$PE, pe$ & Set of processing elements, a processing element\\\hline
%$pe$ & a processing element\\\hline
$K$ & Number of processing elements (e.g., \# of GPUs)\\\hline
$comp(n)$ & Weight of a node $n$\edit{; time required to execute that node on a processing element.} \\\hline
$mem(n)$ & Memory consumption of outputs of a node $n$ \\\hline
$comm(e)$ & Cost of an edge $e$\edit{; time required to communicate the data from the source node (operation) to the destination node (operation).} \\\hline
$sc$ & Secondary cluster, which is a node or a path \\\hline
%a set of one or more nodes where if ${\vert lc \vert > 1}$, then there is a path between each pair of these nodes (all nodes are dependant)
$comm(sc)$ & Total communication cost incurred by all edges that have one end in $sc$\\\hline
$tl(n)$ & Node top level: length of the costliest path between the the source node of the graph and the node $n$, excluding the node $n$. Where the length of a path, is the summation of the computation costs of the nodes on the path and the communication cost of its edges ${\sum_{n \in p}comp(n) + \sum_{e \in p} comm(e)}$ \\\hline
$bl(n)$ & Node bottom level: length of the costliest path between $n$ and the sink node including the node $n$\\\hline
$w\_lvl(n)$ & Node weighted level: $tl(n) + bl(n)$ \\\hline
$span(sc)$ & Time between the expected finish time of the last parent of the first node in a $sc$, and the expected starting time of the first child of the last node in that path. Last and first here mean topologically. \\\hline
$potential(sc)$ & Summation of the weights of all nodes that can be executed within $span(sc)$\\\hline
$st(n)$ & Starting time of node $n$, the time when $n$ is assigned to a ${pe}$ to execute\\\hline
$ft(n)$ & Finish time of node $n$, the time when ${pe}$ is done with executing $n$\\\hline
${M_{cons}(pe, t)}$ & Memory consumed by the processing element at time $t$\\\hline
$M_{pot}(n,t)$ & Memory potential of a node $n$ at time $t$. The summation of the memory occupied by the outputs of $n$'s direct ancestors that are executed before $t$, and for which $n$ is the last direct descendant in its $pe$. Plus $n$'s memory consumption if $st(n) \leq t \leq ft(n)$\\\hline
%the amount of memory reserved due to $n$ and $t$. \didem{due to?}
\end{tabularx}
}
\end{table}

\subsection{Step-1: Partitioning To Minimize Makespan}
This step aims at reducing the makespan of the graph. Before presenting the details of the step, it is important to point  its distinction 
%This step aims at reducing the makespan of the graph 
from both static task scheduling and static mapping.
Unlike scheduling algorithms, we do not specify an order of task execution; we rather focus on spatially allocating the tasks on a set of processors while addressing the locality-parallelism trade-off. The order of execution decision is left to the runtime dynamic scheduler, e.g., TensorFlow scheduler. Unlike static mapping, \algo{} considers the logical and temporal dependencies between the tasks. 

The size, (${\vert V \vert}$), of a large DNNs' computational graph is usually in the order of hundreds of thousands and is projected to grow to millions of operation-nodes~\citep{Rajbhandari2019ZeROMO}.
As a result, efficiency and scalability are essential features of any proposed solution. The multilevel method, the most widely used technique in graph partitioning, addresses the scalability issue by grouping vertices together and dealing with groups of vertices, rather than individual ones~\citep{bichot2011graph}. This grouping reduces the problem size and allows good quality heuristics to be applied within a reasonable time. Inspired by this method, we designed Step-1 of our partitioner in two main stages. 
%that are applied in the following order. 

\subsubsection{{\bf Graph slicing}}
\begin{figure*}[!t]
\footnotesize{
 \begin{center} 
      \includegraphics[width=0.2\textwidth]{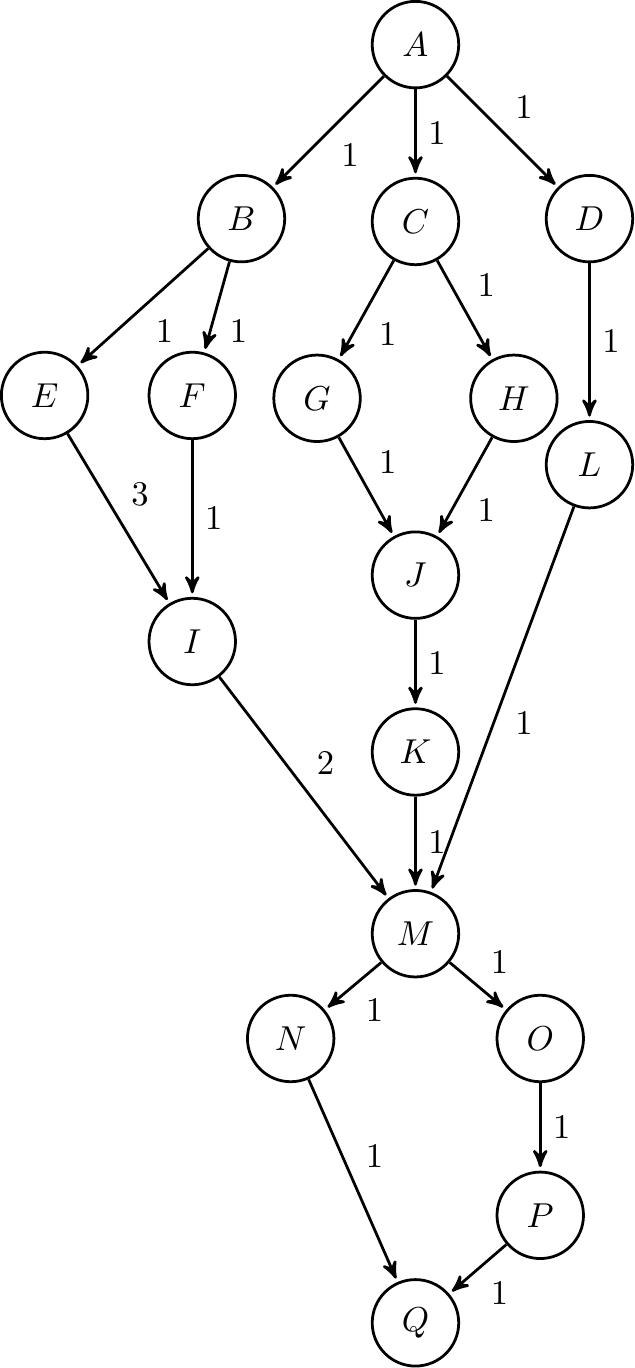}
      \label{fig:graph1}
      \hspace{0.3in}
      %\centering
      \includegraphics[width=0.2\textwidth]{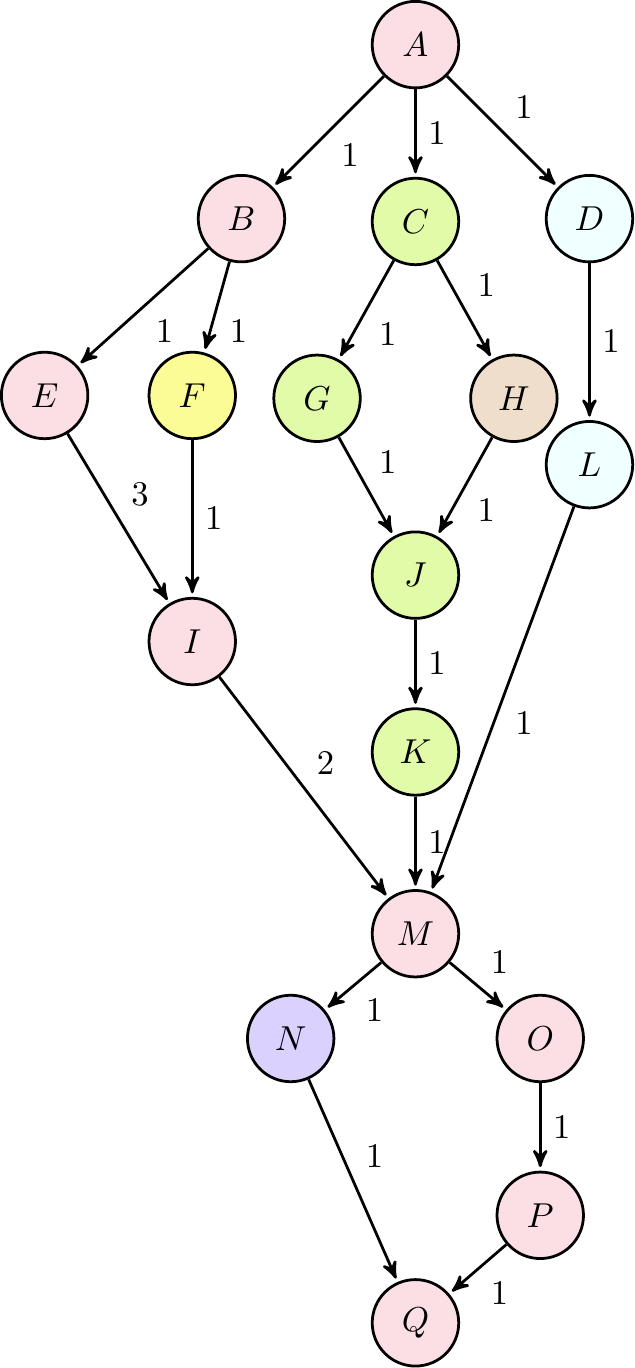}
      \label{fig:graph2} 
      \hspace{0.3in}
      %\centering
      \includegraphics[width=0.2\textwidth]{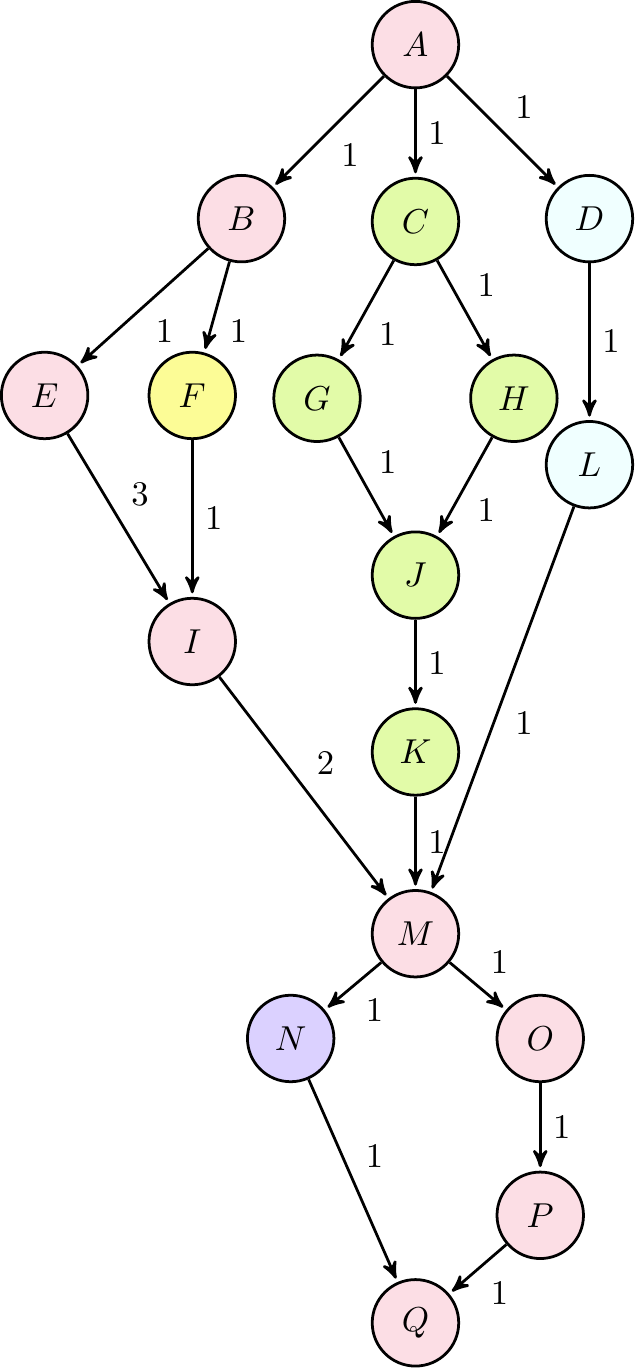}
      \label{fig:graph3}
      \hspace{0.3in}
      %\centering
      \includegraphics[width=0.2\textwidth]{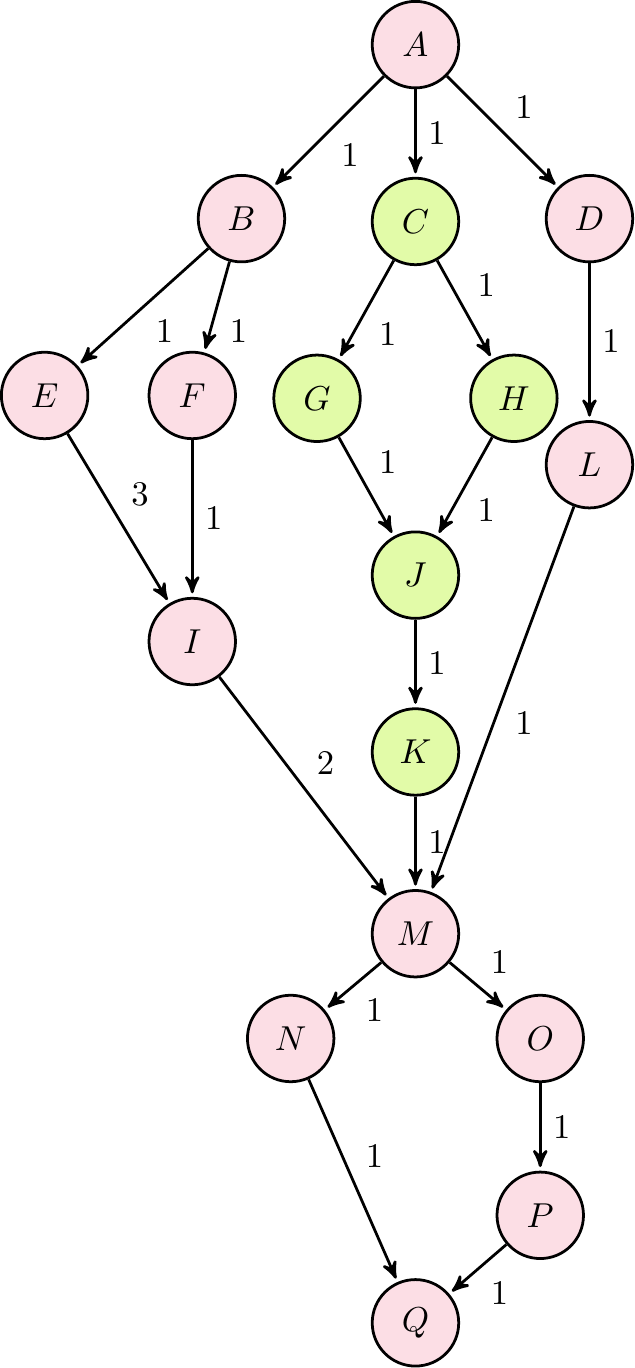}
      \label{fig:graph4}
       \hspace{2.3cm} (a) \hspace{4cm} (b) \hspace{3.8cm} (c) \hspace{4cm}(d) \hspace{1cm}
      \caption{\edit{In the computational graph, edge weights indicate communication costs. All nodes are assumed to have a weight of 1 for simplicity.
      {\bf(a)} Original computational graph. 
      {\bf(b)} Shows the slicing stage when there are two $pe$(s). The obtained clusters are: \{$A, B, E, I, M, O, P, Q$\}, \{$C, G, J, K$\}, \{$H$\}, \{$N$\}, \{$F$\}, \{$D$, $L$\}. First two are primary clusters, the other four are secondaries.
      {\bf(c)} Locality First Lookahead Mapping: Secondary clusters sorted by their criticality, in decreasing order, are \{$H$\}, \{$N$\}, \{$F$\}, \{$D$, $L$\} (note that communications inside a cluster are considered to be zero). \{$H$\} is totally-communicating with a primary and there is enough unmapped work within its span ($L$ and $F$) to cover the temporal/local imbalance. \{$F$\} is totally-communicating with a primary and its mapping does not cause an imbalance (keeping in mind that H is already mapped to the green primary). Both \{$D$, $L$\} and \{$N$\} are totally-communicating as well, but their mapping causes an imbalance and there is no unmapped work left within their spans to cover.
      {\bf(d)} Level Aware Load Balancing: \{$D$, $L$\} and \{$N$\} are mapped according to the criteria in equation  \ref{eq:2}. In th case of \{$N$\} there is a tie (equation  \ref{eq:2} result is 2 for both of the primaries), so it is assigned to the one with which it communicates the most. The makespan is 13.
      }}
      \label{fig:graphs}
      \end{center}}
      \vspace{-0.2in}
\end{figure*}

This stage groups the nodes of the graph into disjoint clusters. It iteratively finds the critical path ($CP$) in the graph and removes $CP$'s nodes and their incident edges from the graph by marking them as visited so that they are not explored in the following iterations. This is repeated $K$ times and the resulting $K$ many $CP$s are called \emph{primary clusters}, which are the initial partitions assigned to different processing elements. Hence, the terms primary cluster and ${pe}$ are going to be used interchangeably.
After finding those primary clusters, the remaining nodes are grouped into {\em secondary clusters}. A secondary cluster, which is a linear cluster~\citep{sinnen2007task}, is either a single node or a path. All the secondary clusters are identified and tagged until there is no node left on the graph that is not part of any cluster. Figure \ref{fig:graphs}(b) shows an example.

\begin{algorithm}[!t]
\footnotesize{
	\caption{Graph Slicing} 
	\label{alg1}
	\hspace*{\algorithmicindent} \textbf{In :} K, Graph G \\
 \hspace*{\algorithmicindent} \textbf{Out:} pri\_clusters[ ], sec\_clusters[ ] \Comment{{\em initially empty}}
	\begin{algorithmic}[1]
%	    \State ${pri\_grps \gets \phi}$
%	    \State ${sec\_clusters \gets \phi}$$
		\State $j \gets 1 $
		\State ${w\_lvls \gets}$ compute\_weighted\_levels(${G}$)
		\While {${G \neq \phi}$}
		    \State ${heaviest\_path \gets}$ find\_heaviest\_path(${G, w\_lvls}$)
		    %\\  \hskip\algorithmicindent \hskip\algorithmicindent find\_heaviest\_path(${G, w\_lvls}$)
		    \If{${j \leq K}$}
		    \State $pri\_clusters[j] \gets heaviest\_path$
		    \State ${w\_lvls \gets}$ compute\_weighted\_levels(${G}$)
		    \Else
		    \State $sec\_clusters[j-K] \gets heaviest\_path$
		    \EndIf
		    \State ${G \gets G - \{heaviest\_path\}}$
		    \State ${j \gets j + 1}$
		\EndWhile
	\end{algorithmic}
	}
\end{algorithm}
\setlength{\textfloatsep}{0pt}

Algorithm~\ref{alg1} shows the pseudo-code of the graph slicing, which takes device count $K$ and graph $G$ as inputs and outputs primary and secondary clusters. Line 2 computes the weighted level ($w\_lvl(n)$) for all the nodes in the graph. The heaviest path, (Line 4), is the $CP$ when $w\_lvl(n)$ are recalculated.  Finding the heaviest path is done by traversing the graph using the computed ${w\_lvls}$ as priorities until reaching a dead-end. After forming a $CP$, it is added to the primary clusters and its nodes and edges are removed from the graph (Line 11). Unlike linear clustering~\citep{kim1988general}, we obtain only $K$ many $CP$s, then we stop recalculating $w\_lvl(n)$ for the secondary clusters since computing weighted levels is expensive.  In section \ref{sec:lc_cp} we demonstrate that avoiding this expensive computation does not harm the quality of the results. 
~\edit{When weighted levels are not recalculated, $find\_heaviest\_path$ may not return a $CP$, rather returns a path of a heavy cost. Thus the term \emph{heaviest path} refers to the heaviest path from the slicing algorithm perspective, which is not necessarily the actual critical path. } 
%This aims at capturing dependent and heavily-communicating nodes in one cluster to increase locality. 
If a path could not be obtained, it returns a single node.
%~\edit{We use the term \emph{heaviest path} to refer to the heaviest path from the slicing algorithm perspective, which is not necessarily the actual heaviest/ critical path(the last can be obtained if the weighted levels are recalculated)}.

{\bf Complexity:} The most expensive part of Algorithm~\ref{alg1} is computing weighted levels for all the nodes. 
This operation performs a variant of topological sorting and has time complexity of ${O(\vert V \vert + \vert E \vert)}$~\citep{sinnen2007task}. It is done $K$ times, resulting in an overall complexity of ${O(K(\vert V \vert + \vert E \vert))}$. In linear clustering that would cost ${O(\vert V \vert(\vert E \vert + \vert V \vert))}$~\citep{wang2018list}.  Given that the priorities are already specified, finding the paths costs ${O(\vert E  \vert)}$\footnote{\edit{The task graph is assumed to be connected graph, which means it has at least $n-1$ edges}}. This is because each node is visited at most once, since the nodes are removed from the graph (marked as visited) once they join a path. From each node deciding the next to visit entails picking its highest priority neighbor, which requires checking its adjacent nodes' priorities. Since the edges are removed from the graph with their incident vertices as well, no edge will be visited twice. Hence, overall there are ${O( \vert E  \vert)}$ steps. It is important to note that all the graphs we experimented are sparse, having ${ \vert E  \vert <  \vert V  \vert log( \vert V  \vert )}$. %, the rest of the analyses adopts this assumption. 

\subsubsection{{\bf Mapping}}
This stage attaches the secondary clusters to the primaries with the goal of obtaining a partition with minimal makespan by addressing the locality parallelism trade-off. This process is referred to as {\em cluster mapping} in the scheduling literature. There are mapping heuristics such as wrap cluster merging~\citep{yang1993scheduling}, list scheduling based cluster assignment ~\citep{sarkar1988partitioning}, and Guided Load Balancing (GLB)~\citep{radulescu1998glb}. 
%The first two alternatives have their own shortcomings~\citep{radulescu1998glb}.
In a comprehensive evaluation of scheduling and cluster merging algorithms in~\citep{wang2018list}, GLB is shown to produce the best result. However, GLB assumes that its preceding clustering step has eliminated the largest communication delays. As a result, the communication delays are not considered for cluster mapping~\citep{radulescu1998glb}. Ignoring communication cost results in a low-quality mapping when the graph becomes very large. 
%and the amount of communication increases. 
Even if each inter-cluster communication is small, the cumulative effect becomes considerable. In addition, the load balancing is global rather than time dependent (temporal). This balancing is not suitable especially for graphs with frequent forks and joins (e.g., DNN graphs), where the local and the global loads become more uncorrelated. 
\edit{We propose a novel time-efficient heuristic called {\em Locality-First Lookahead and Level-Aware Mapping (LFLAM)}. LFLAM considers both critical-communication minimization and the temporal load balancing. 

\paragraph{\bf Locality-First Lookahead Mapping}
First, we perform locality-first lookahead mapping, which assigns a secondary cluster $sc$ to the primary with which it communicates the most as long as there is unmapped work (other tasks ahead) within ${span(sc)}$ that can cover the temporal load imbalance caused by that mapping step. The intuition is that if both the balance and the locality are considered in the cost function at the beginning of the mapping, the locality might be sacrificed in some cases to account for balance. In such a case, an additional refinement step would be needed to improve locality. Such a refinement step would swap some of the already mapped clusters so that the locality is regained without hurting the load balance. However, optimizing for the locality first, knowing that there is a sufficient amount of tasks ahead to balance again, diminishes the need for such a refinement step. On the other hand, overprioritizing locality may harm parallelism. Hence, it is crucial to decide which paths should be targeted. 
Moreover, some communications are less important to eliminate than others. For example, a communication edge between two nodes that have enough amount of work to hide (e.g., $(A, D)\ and\ (L, M)$ in Figure \ref{fig:graphs} b) is not as important as communication that can be a pure delay (e.g., $(H, J)$).

%that can cover the temporal load imbalance cased by that mapping step. 
%The intuition behind prioritizing locality is the following: in a certain section of the graph, given a set of equally-balanced mappings, the one with the highest locality is preferred. If both the balance and the locality are considered in the cost function at the beginning of the mapping, the locality will be sacrificed in some cases to account for the balance. 

%To improve the locality, an additional refinement step will be needed to swap some of the mapped clusters so that the locality is regained while causing a negligible load-imbalance compared with the gained locality or none at all. However, optimizing for the locality first, knowing that there is a sufficient amount of tasks ahead to balance again, diminishes the need for such a refinement step. 

%However, over prioritizing locality may harm parallelism. Hence, it is crucial to decide to what extent should the locality be prioritized, or which paths should be targeted. Moreover, some communications are less important to eliminate than others. For example a communication edge between two nodes that have enough amount of work to hide -like $(A, D)\ and\ (L, M)$ in Figure \ref{fig:graphs} b- is not as important as a communication that can be a pure delay -like (H, J) in the same figure-. 

%{\bf Target Paths for Locality:} 
We decide which paths to be targeted based on $CCR$ (total communication cost to total computation cost ratio) of the graph~\cite{ozkaya2019scalable,sinnen2007task}. Higher $CCR$ means more communication compared to computation, hence locality is more important.  In the literature~\cite{sinnen2007task}, graphs with $CCR \geq 10$ are considered as highly communicating. Hence we use the $CCR$ value as a threshold to decide to what extent to prioritize the locality. 
We divide the secondary clusters into two groups: totally-communicating and maximally-communicating clusters. Totally-communicating clusters are the ones whose all the incoming and outgoing communications are with only one of the primaries.
A maximally-communicating cluster is the one whose communication with one of the primaries is greater than its all communication/$K$. Note that this does not always hold since an $sc$ can be communicating with other secondaries not only with primaries. 
For graphs which have equal or higher $CCR$ value than the threshold, clusters from both totally-communicating and maximally-communicating are considered because we need to save as much communication as possible. Otherwise only totally-communicating clusters are considered. These clusters should be added to their primaries even when $CCR$ is low as long as there is unmapped work within ${span(sc)}$ that can cover the possible temporal load imbalance because such mapping is a pure gain as it would cancel all their communications.

%\begin{equation}
%\label{eq:1}
%$\exists pe \in PE \sum_{ (u, v), u \in sc, v \in pe} comm(u, v)\  + \sum_{ (v, u), u \in sc, %v \in pe} comm(v, u) \geq comm(sc) / K$ 
%\setlength{\textfloatsep}{0pt}
%\end{equation}

%(note that this does not always hold since a $sc$ can be communicating with other secondaries not only the primaries). 
%For graphs which have equal or higher $CCR$s, clusters from both totally-communicating and maximally-communicating clusters are considered because we need to save as much communication as possible. Otherwise only totally-communicating clusters are considered, these should be added to their primaries even when $CCR$ is low, as long as the assumption that there is unmapped work within ${span(sc)}$ that can cover the possible temporal load imbalance holds, because such mapping cancels all their communications.

%
We decide the most important communication edges among clusters based on their {\em criticality}. The criticality of a linear cluster $lc$ is the length of the longest path going from the graph source to its sink and completely overlapping with $lc$. This is equivalent to the $w_{lvl}(n)~\forall n \in lc$, where $w_{lvl}(n)$ is recalculated by setting communications within $lc$s to zeros after the slicing stage of \algo. 
The clusters are sorted by their criticality in a non-increasing order. 
Then the desired clusters, depending on the $CCR$ threshold, are traversed and any secondary cluster is mapped to its most communicating primary if any of three conditions holds: (a) $lc$ has enough unmapped work within its span to cover the local load imbalance of its mapping if any is caused by the mapping, (b) if no local load imbalance occurs or the local balance is improved, and (c) $lc$ communication with the target primary is larger than its weight, the work of the primary within its span, and the unmapped work in its span. In the case of condition (c) if mapping were not done, the communication would dominate potentially creating a new longer critical path.

\paragraph{\bf  Level-aware Load Balancing} 
Next, level-aware load balancing is applied to map the clusters that were not mapped in the previous step. Temporal balance of the loads is achieved by considering the workload of every ${pe}$ within ${span(sc)}$, where ${sc}$ is the secondary cluster that is going to be mapped to one of the primary clusters. ${sc}$ is mapped to a ${pe}$ that has the minimal computational load within the ${span(sc)}$, and minimizes the incurred communication with the other processing elements. Equation~\eqref{eq:2} shows the selection criteria. In case of ties, we assign $sc$ to the ${pe}$ which has the highest communication value with it.
}
\begin{equation}
\label{eq:2}
%\begin{split}
\min_{pe \in PE}\Big(\
    \sum_{\mathclap{\substack{ n \in pe, \\
                               tl(n) \in span(sc)}}} 
       comp(n) + 
    \sum_{{\mathclap{\substack{ (n,u) \in E, \\
                                n \in \{PE\} - pe, \\ 
                                u \in sc}}}} 
       comm(n,u) + 
     \sum_{{\mathclap{\substack{(u,n) \in E, \\ 
                                n \in \{PE\} - pe, \\ 
                                u \in sc}}}} 
       comm(u,n)\Big)
%\end{split}
\setlength{\textfloatsep}{0pt}
\end{equation}
    
\edit{Algorithm ~\ref{alg2} shows a high level description of mapping. Lines 1-7 perform the Locality-First Lookahead Mapping.~$locality\_first\_lookahead\_mapping$ takes the secondary clusters as a parameter and decides whether to map the maximally-communicating clusters or not depending on the $CCR$ value. It then maps the clusters based on the three conditions we explained in the previous paragraph and returns the number of the mapped clusters. Line 6 could be repeated as long as at least one path has been mapped since the mapping changes the communication. This possibly leads to the formation of new totally-communicating or maximally-communicating paths. We choose to repeat this step at most $log(|V|)$ times to maintain low complexity.
 The while loop (in Line 10) applies the level-aware load balancing. The most time consuming part in this algorithm is to calculate the work within the span of the target secondary cluster $sc$ both as a part of $locality\_first\_lookahead\_mapping$ and on (Line 12) in each of the primary clusters.} We model this part as a problem of frequent range queries with updates. More specifically, we need to find the sum of the weights of the nodes whose levels fall in the span, and upon merging, the weights of those levels are updated. We use binary-indexed-trees~\citep{fenwick1994new} as a data structure, where the tree nodes store the weights per level. This data structure allows logarithmic range summation and value updates. Line 13 calculates the cost of communication between the secondary cluster $sc$ in each of the primary clusters. Line 14 performs the selection criteria defined in Equation~\eqref{eq:2} to select the best primary cluster to merge the $sc$ with.

{\bf Complexity:} 
Before starting LFLAM, we sort the clusters by their weights, this has an upper bound of ${O(\vert V \vert * log\vert V \vert)}$ since the number of clusters is upper-bounded by the number of nodes. 
Since the clusters are disjoint paths or singular nodes, and have no common nodes, the total number of  the update operations is bounded by ${\vert V \vert}$. The number of range summation queries is bounded by the number of the paths which is again bounded by ${\vert V \vert}$. The cost of either of the operations is logarithmic in the number of the levels. The number of the levels is ${\leq \vert V \vert}$, so we end up with ${O(\vert V \vert * log\vert V \vert)}$.  
\edit{The Locality-First Lookahead Mapping is repeated $log(\vert V \vert)$ times. %leading to an overall complexity of ${O(\vert V \vert * log^2\vert V \vert)}$.}
Hence, the overall complexity of the mapping stage is ${O(\vert V \vert * log^2\vert V \vert)}$.}

\algdef{SE}[DOWHILE]{Do}{doWhile}{\algorithmicdo}[1]{\algorithmicwhile\ #1}%

\begin{algorithm}[t]
\edit{
\footnotesize{
	\caption{LFLAM Mapping} 
	\label{alg2}
	 \hspace*{\algorithmicindent} \textbf{In:} pri\_clusters[ ]\\
     \hspace*{\algorithmicindent} \textbf{In:} sec\_clusters[ ] 
	\begin{algorithmic}[1]
		\State ${w\_lvls \gets}$ compute\_weighted\_levels(${G}$)\\
		sort(sec\_clusters, criticality\_sorting\_criteria)
		\State num\_of\_mapped\_lcs $\gets 0$
		\State iteration $\gets 0$
		\Do
        \State num\_of\_mapped\_lcs $\gets$ locality\_first\_lookahead\_mapping(sec\_clusters)
        \doWhile{num\_of\_mapped\_lcs $> 0$ and iteration $\geq log(|V|)$} \\
		\State $comps [\ ]\gets \phi$, $comms [\ ]\gets \phi$
		\While {${sec\_clusters \neq \phi}$}
		    \State $sc \gets$ remove\_next\_secondary($sec\_clusters$)
		    \State ${comps \gets}$ calc\_work\_at\_span(${span(sc)}, pri\_clusters$)
		    
		    \State ${comms \gets}$ calc\_comms\_with($sc$, $pri\_clusters$)
		    \State ${target\_pri \gets}$  find\_minimal(${comps, comms}$)
		    \State ${target\_pri \gets target\_pri + \{sc\}}$
			% \State main\_path\_of\_minimal(${comps,}{comms[other\_main\_groups]}$)
			%\hskip\algorithmicindent \hskip\algorithmicindent 
		\EndWhile
		
	\end{algorithmic} 
	}}
\end{algorithm}
\setlength{\textfloatsep}{1pt}

\begin{comment}
\begin{figure}
\begin{center}
\includegraphics[scale=0.7]{figs/orig.PNG}
\includegraphics[scale=0.7]{figs/sliced.PNG}
\caption{Slicing the computational graph. The paths are found in the following order:\\
${['s', 'n2', '4', 'n6', 'n8', 'n10', 'k'],\ ['n3', 'n5'],\ ['n9'], ['n1'],\ ['n7']}$\\s
Note: the numbers inside the nodes indicate the computation time, and the number inside the edges represent the communication time.\wahib{refer to the figure in the text}}
\label{fig:slicing}
\end{center}
\end{figure}

\begin{figure}
\includegraphics[scale=0.7]{figs/init_merged.PNG}\hspace{1cm}
\includegraphics[scale=0.7]{figs/lalb.PNG}\hspace{1cm}
\includegraphics[scale=0.7]{figs/glb.PNG}\\
(a) Initial Merging \hspace{1.5cm} (b) LALB \hspace{1.5cm} (c) GLB
\begin{center}
\caption{Initial Merging and comparison between LALB and GLB: (a) path ${['n9']}$ is merged since it has a communication of $5$ units that cannot be hidden by the work within its span. (b) and (c) show LALB and GLB applied after (a) (assuming we have ${2 pe's}$). The makespan of the LALB output is ${13}$ units, while GLB is ${15 -15\% slowdown-}$. Ignoring the temporal aspect of load balancing causes GLB to make an incorrect decision for ${n3}$. It assigns it to the less loaded ${pe}$, yet that ${pe}$ has more work within the ${span(['9'])}$. This incorrect assignment patterns happens frequently in DNN DAGs due to the large number of levels in the graph, as later shown in the results section.\wahib{refer to the figure in the text}}
\end{center}
\end{figure}
\end{comment}

\subsection{Step 2: Handling Memory Overflow}
\label{mem_const}
Similar to Step-1, we handle the memory constraints statically ahead of time for two main reasons: (a) to avoid any runtime overhead,
%, the allocation is decided once and can be used as long as the model parameters that affect the memory consumption are not changed;
and (b) to reduce the chance of conflicting with other runtime optimizations. 
%Optimizing memory consumption of production DL frameworks is an area of an ongoing development and the memory management modules of those frameworks go through continuous modifications. 
%As a result, the memory management modules of those frameworks are prone to continuous modifications. 
\edit{We propose two memory heuristics both of which could be seamlessly used with any optimization policies provided by the DL frameworks. First heuristic is emulation-based and guarantees to meet the memory capacity constraint. The second is a practical method that aims to balance memory consumption on devices.  }
%\algo{} results in an allocation that meets an upper bound of memory consumption, and a dynamic policy could use that result as an initial point to further optimize at runtime. Step-2 is further divided into three stages; a scheduler emulator, memory consumption tracking, and overflow handling. 

\subsubsection{Memory Heuristic I} 
This heuristic has three components; scheduler emulator, memory consumption tracker, and overflow handler.
\paragraph{\bf Scheduler Emulator} To address the memory consumption statically, temporal modeling of the allocation and deallocation patterns is required. Such modeling necessitates knowledge about scheduling in the DL framework to estimate when an operation is going to start and finish execution. Consequently, when the memory allocated for the operation inputs is released and when a new memory is allocated to hold the operation outputs. To estimate those values, we emulate TensorFlow scheduler. TensorFlow scheduler maintains a ready queue that is initially filled with nodes with no ancestors. Each node in the graph has an in-degree representing the number of nodes it depends on. The nodes are executed in FIFO order. Once a node is executed, the in-degrees  of its children are decremented by one. Any node having an in-degree of zero will be pushed to the queue. Using the per-node running times and communication sizes collected from profiling, we emulate this behaviour to get the expected start- and end-times of the operations under a certain partitioning. %, whose implementation will be explained in Section \ref{sec:imp}.
%\fareed{We are done with the emulator up to here. Emulator is just to tell the expected starting time, that's it. It has nothing to do with nodes types the coming is our modeling and may be under memory tracking as now}

{\bf Complexity:} The scheduler emulator estimates the starting time ${st(n)}$ and finishing time ${ft(n)}$ of the nodes in the graph. The emulation has a time complexity of ${O(\vert V \vert+ \vert E \vert)}$ as the nodes are visited and on each visited-nodes, the in-degrees of its direct descendants are decreased.

\paragraph{\bf Memory Consumption Tracker}
\label{sec:trackingmem}
In TensorFlow, from the memory consumption perspective, operation-nodes broadly fall into three main categories. First, operations of which the data survives across the iterations~\citep{abadi2016tensorflow} and we refer to them as {\em residual nodes} (${res\_ns}$). Second, special operations that mutate the referenced tensor, of the first type, we refer to them as {\em reference nodes} (${ref\_ns}$). Those operations do not reserve any additional memory. However, they are co-located with the variables that they are mutating and must be moved together with their referred to variable nodes. Third, operations that require additional memory proportional to their output size and we call them {\em normal nodes} (${nor\_ns}$). Memory for the output of these nodes is allocated upon scheduling and released once all their direct descendants are executed. 

To create a functional cost model, our memory consumption estimation takes into account the scheduler, node types, and profiled per-node memory consumption.  One might think that profiling solely peak 
memory footprints would be sufficient to predict the overflows. However, to handle an overflow, nodes have to be moved between partitions and that in turn changes the schedule and the memory consumption as a function of time. 
Our cost model takes this dynamic behavior into account and models the interplay between the scheduler and memory usage.

Equation\eqref{eq:3} defines the memory consumption of a ${pe}$ at time $t$, $M_{cons}(pe, t)$. 
The first term is the memory consumption of the ${res\_ns}$ assigned to that ${pe}$. 
The second term indicates the memory consumption of the normal nodes that have started on that ${pe}$ at $\leq {t}$ and are being executed at ${t}$. 
The third indicates the nodes that have descendants assigned to that ${pe}$ and the descendants' expected starting time is $\geq{t}$, and those nodes have finished at $\leq {t}$ at any processing element except ${pe}$, or are non-residual that  have finished at ${\leq t}$ on that ${pe}$.

\begin{equation}
%\begin{aligned}
\label{eq:3}
%\begin{split}
M_{cons} (pe, t) = 
    \sum_{{\mathclap{\substack{n \in pe, \\
                                \ n \in res\_ns}}}} 
    mem(n) 
    +
    \sum_{{\mathclap{\substack{n \in pe, \\
                                \ n \in nor\_ns,\\
                                st(n) \leq t \leq ft(n) }}}} 
    mem(n) 
    + 
    \sum_{{\mathclap{\substack{n \notin (pe \cap res\_ns), \\
    ft(n) \leq t,\\ \exists (n, u) \in E: st(u) \geq t, u \in pe
                              }}}} 
    mem(n)
%\end{split}
%\end{aligned}
\end{equation}
%\vspace{0.1in}

%from Equation \eqref{eq:3}, it can be inferred that there are three cases for which the memory is reserved on a ${pe}$ at time ${t}$. For a node $n$ on ${pe}$, (i) $n$ is a ${res\_ns}$, or (ii) $n$ is a ${nor\_ns}$ and is going to be executed at time ${t}$, or (iii) $n$ ${\notin res\_ns}$ and is going to be executed after ${t}$ and has a predecessor $u$ ${\notin res\_ns}$}  that is going to be executed before ${t}$.

%\didem{will continue from here after playing with Poyraz.}
%The change in memory consumption is triggered by a node execution, thus we care about the starting times of the nodes. Once a node starts executing, new memory needs to be allocated and that may cause an overflow. So 
%\edit{Following this equation and using the times driven from the scheduler emulator, we map the per-node memory consumption values collected using TensorFlow profiler to a memory consumption. A valid question is why not profiling the overall consumption in the first place? The answer is that it is not enough to solve the overflow. While addressing the memory constraints we move nodes which in turn changes the schedule and the overall consumption which we need to keep track of. Thus it is necessary to have that modeling.}  

The overall memory consumption needs to be estimated for each node (${\vert V \vert}$ time points) because the change in memory consumption is triggered by node executions. Once a node starts executing, new memory space needs to be allocated and that may cause an overflow. Estimating memory consumption is done by visiting all the nodes in the graph in the order of their estimated starting times, which is obtained from the scheduler emulator, and keeping track of the accumulated memory consumption. In the same pass, the memory potential values of the nodes ($M_{pot}$ in Table \ref{tab:terms}) are obtained. A node's memory consumption is added to the cumulative value once it is visited, and subtracted after its last descendent in a certain ${pe}$ is visited unless it is a ${res\_ns}$. 

{\bf Complexity:} Tracking the memory consumption requires ${O(\vert V \vert)}$ time since it is done in one pass over the graph nodes while keeping the cumulative values and calculating the potentials. This is given that the node last descendant assigned to each processing element is knows; which is collected while emulating the scheduler and hence its complexity is implicitly included in the scheduler emulator part. 

\paragraph{\bf Overflow Handler}
\label{sec:overflow}
After estimating the memory consumption, we traverse the graph starting from the sink and keep the nodes 
%that are normal or residual nodes \didem{reference?}
%that fall in one of the three aforementioned categories 
in a heap data structure, namely ${nodes\_heap}$. When the memory consumed exceeds the limit, we deal with the overflow as a 0-1 min-knapsack problem~\citep{csirik1991heuristics}. The min-knapsack problem is formulated as follows; given $N$ pairs of positive integers (${c_j}$, ${a_j}$) and a positive integer ${O}$, find ${x_1}$, ${x_2}$, ..., ${x_N}$ so as to:
%\vspace{-0.05in}
\begin{equation}
%\begin{aligned}
\label{eq:4}
%\begin{split}
minimize \sum_{j=1}^{N} c_j x_j  \quad \quad 
s.t.\ \sum_{j=1}^{N}a_j x_j\geq O \ and \ x_j \in \{0, 1\}
%\end{split}
%\end{aligned}
\end{equation}

In our case, ${O}$ represents the amount of memory {\em overflow}, and ${a_j}$ represents ${M_{pot}(n, t)}$.
%\didem{it was $mempot(n_j)$ before, not clear? what is $n$ here?}\fareed{discuss}. 
For the cost $cj$, we use the summation of the node computation cost and the communication with its direct ancestors and descendants located on the same ${pe}$, as shown in Equation\eqref{eq:5}, which defines {\em move\_cost}. 

\begin{equation}
\label{eq:5}
comp(n)+\sum_{\forall u,v \in pe(n)}(\sum_{u:(u,n) \in G} comm(u,n) + \sum_{v:(n,v) \in G} comm(n,v) )
%\end{split}
\end{equation}

The idea behind {\em move\_cost} is that when a node is moved from a ${pe}$ to another, it incurs a computational load imbalance proportional to its weight and extra communication proportional to its communication with the nodes assigned to the same ${pe}$. Our goal is to find a set of operation-nodes that the summation of their memory consumption potentials at the overflow time is ${\geq}$ {\em overflow} when their total movement cost is minimized. The {\em movement criteria} is to pick the node that has the lowest {\em move\_cost}$/ M_{pot}(n, t)$. In other words we choose the node that alleviate the overflow while incurring the least amount of communication and computation imbalance. 
%\didem{fix M(n,t)s}

The ${nodes\_heap}$ is a min heap in which the {\em movement criteria} is the ordering key. To avoid choosing a node that has a low {\em movement criteria} but high {\em move\_cost}, each node for which the ${M_{pot}(n, t) >}$ {\em overflow} is inserted in another heap at which the sorting key is {\em move\_cost}. When selecting, the top node is removed from both heaps and the one with the least {\em move\_cost} is chosen, and the other is returned to its heap.  The selected node is moved to another $pe$ if the target $pe$ has sufficient memory to accommodate that node memory potential. Otherwise, the node is not considered again and another node is picked from the heap. The algorithm terminates when either the overflow is eliminated or we run out of nodes without addressing it.

 {\bf Complexity:} We solve the knapsack greedily as the dynamic programming based solution complexity is impractical. When an overflow is detected, we pick the nodes from the heaps in a logarithmic time. 
 Any node that is moved to another partition is guaranteed not to be moved again since it is moved only if the destination $pe$ can accommodate it, meaning that it can neither cause nor solve an overflow on that $pe$. As a result, there is no repetition and a node can enter or exit the heap once, resulting in $O(\vert V \vert *log \vert V \vert)$. When a node is moved, the new potentials and memory consumption need to be recalculated (${O(\vert V \vert)}$). It may happen at most $\vert V \vert$ times. Overall the complexity is $O(\vert V \vert^2)$. However, 
 %as shown in Section \ref{sec:results} 
 this upper bound is much larger than the practical one as the number of nodes to be moved is usually much less than $\vert V \vert$.

\edit{
\subsubsection{Memory Heuristic-II}
\label{sec:heu2}
%Heuristic-I guarantees to meet the memory constraints of the devices.  
Heuristic-II balances the memory consumption among devices but theoretically does not guarantee meeting the memory constraint. 
%The disadvantage of this heuristic is that, theoretically, it does not guarantee meeting the constraints. 
Its advantages are (a) it does not require the emulation of the DL scheduler, and (b) it is practically efficient  as it permits training all the models trainable using Heuristic-I in our experiment set. 

This heuristic is based on the observation that most of the long-living memory reserved throughout an iteration (one execution of the graph), is either for a resident node holding parameters, 
or a normal node holding output that will be fed to a distant node in the critical path.
We apply two strategies  to alleviate any possible high imbalance in memory consumption between the partitions. The first balances the residual and non-residual net memory consumption ratios across the partitions. 
Since the memory for residual nodes, which is allocated for the model parameters, is reserved and not freed until the training ends, it forms a permanent memory pressure. Moreover, most of these nodes are not critical nodes. Only the ones used in the first few layers should be provided quickly, meaning that their communications are critical. 
The rest, which are the vast majority, can be communicated by the time they are needed, which makes them good candidates to move. We move these nodes from the most loaded partitions to the least loaded ones until the summation of ratio of the residual node memories in a partition to the overall memory for residual nodes and the ratio of the normal node memories in a partition to the overall memory for normal nodes are as equivalent as possible. 
For example, if there are two partitions, where partition $1$ holds $70\%$  of the normal nodes' memory and $60\%$ of the residual nodes' memory (partition $2$ holds $30\%$ and $40\%$, respectively),
%and residual nodes having $60\%$ of the residual nodes memories 
then the result of applying the strategy would be partition $1$ having $70\%$ of the normal and $30\%$ of the residual nodes' memory, and partition $2$ having $30\%$ and $70\%$, respectively. In other words, the residual nodes are moved until Equation~\ref{eq:6} holds or there are no more residual nodes to move.

\begin{equation}
%\begin{aligned}
\label{eq:6}
\begin{split}
\forall pe \in PE,\ K / 2 \ (
    \sum_{{\mathclap{\substack{n \in pe, \\
                                \ n \in res\_ns}}}}  
    mem(n) \ / \sum_{{\mathclap{\substack{
                                n \in res\_ns}}}} mem(n)
    \ + \\
    \sum_{{\mathclap{\substack{n \in pe, \\
                                \ n \in nor\_ns}}}} 
    mem(n) \ / \sum_{{\mathclap{\substack{n \in nor\_ns}}}} 
    mem(n) ) \geq 1
\end{split}
%\end{aligned}
\end{equation}
%\vspace{0.1in}

In some cases moving the residual nodes is not enough. This happens when the graph is very thin, hence most of the nodes are located on the critical path (the first path of the $K$ paths discovered by the slicing algorithm). 
This case results in a highly unbalanced partitioning because a balanced partitioning, in this case, would require splitting the critical path which is not desired as it means longer execution time. However, if the memory of a single device is not sufficient to hold the peak memory that is required to execute the critical path then the path has to be divided to be able to run the graph. So, our second strategy detects if there is a high imbalance between the primaries and gives contiguous chunks of the longest to the shortest -one chunk for each of the shortest- to balance the memory requirements. This is performed right after slicing but before mapping stages so that the Locality-First Lookahead mapping guarantees the highest level of locality around the reassigned chunks and the level-aware load balancing balances the secondaries accordingly. 
Moving contiguous chunks of the critical path does not cause a considerable performance degradation in the case of our large and thin graphs since the critical path of a thin DNN graph contains thousands to tens of thousands of nodes and introducing few communications has a negligible effect. Without emulation or knowing the schedule, it is not possible to know the memory consumption pattern and the bottlenecks. As a result, heuristic-II should be enabled or disabled by the user, as the user may want to train with a certain batch size permitted by the partitioning without applying a memory heuristic. For example, for the models that fit into a single device memory using a small batch size, only partitioning the model is sufficient for a linear scaling of the batch size.  

{\bf Complexity:} This heuristic requires simply moving nodes among partitions. Since the movement is unidirectional, from the most loaded to the least, each node is moved at most once. Hence the total number of moves is bounded by the number of nodes resulting in a time complexity of ${O(\vert V \vert+ \vert E \vert)}$.
}

\section{Implementation}
\label{sec:imp}
%\wahib{The implementation in TensorFlow is discussed here. This section would be informative for those who implement frameworks, however if space becomes an issue, we could shrink it to just give an overview of the implementation and emphasize the effort that was required. Then we could mention that we intend to release a separate and detailed tech report for the implementation+code. If there is space to expand, then we could move it to be a stand-alone section after Section~\ref{sec:pardnn}}
%The memory capacity of Nvidia GPUs can be collected using the $nvidia$-$smi$ command. The practical interconnection bandwidth and latency can be collected using CUDA toolkit ${p2pBandwidthLatencyTest}$. 
Our algorithm takes as an input the device count, their memory capacities, the interconnection bandwidth and latency between them, the model computational graph, profiling data, operations metadata. The profiling data contain execution time measurements and the size of the output of each operation-node. The operation metadata contain the operation types (section~\ref{sec:trackingmem}).
TensorFlow standard APIs provide the profiling information including per-node time, memory consumption, and communication sizes at the granularity of graph nodes for regular as well as user-defined operators. 
%User-defined operators are supported as these operators have to be registered through the TensorFlow APIs as any regular operation. 
%. We do not rely on the specific operations types; this data is merely used to identify the node types. 
%All this data can be collected using standard APIs provided by TensorFlow.~\edit{All the profilings are collected at the graph-node level. Tensorflow profiling API's provide the running time, memory consumption, and communication size values for each of the operations in the graph.

To estimate the memory consumption, we implemented an emulator of TensorFlow's scheduler described in~\cite{abadi2016tensorflow}. It is important to note that if \algo{} with memory heuristic-I is intended to be used with another DL framework, another emulator can be written to emulate its scheduler, if needed, without modifying our partitioning algorithm.
%It is important to note that our memory handling part is loosely-coupled to this emulator, and independent of its internals. It just takes the sequence in which the nodes are expected to be executed. If intended to be used with another framework, another procedure can be written to emulate its scheduler ,if needed, without touching our algorithm. 
%From the memory lifetime point of view, TensorFlow operation-nodes fall into three main categories, {\em residual, reference, and normal} nodes, as described in Section \ref{sec:trackingmem}.  
%Operations of which the data survives across the iterations ~\cite{abadi2016tensorflow}, namely {\em Variable} nodes. We call them {\em residual nodes} (${res\_ns}$).
%(e.g., a variable which is a special kind of operation that returns a handle to a persistent mutable tensor~\cite{abadi2016tensorflow}).  
%(e.g., Assign and AssignAdd (equivalent to +=))
%Second, special operations that mutate the referenced tensor, of the first type, we refer to as {\em reference nodes} (${ref\_ns}$). Those operations do not reserve any additional memory. However, reference operations are 'co-located' by TensorFlow with the variables that they are mutating, and have to be moved together in case the variable they are referring to is moved.
%Third, operations that require additional memory proportional to their output size. Those operations, we call {\em normal nodes} (${nor\_ns}$), memory for these nodes outputs is allocated upon scheduling and released once all their direct descendants are executed. 
When handling memory constraints there is a trade-off between the overhead and the accuracy; static handling prioritizes overhead reduction over accuracy while dynamic handling targets the opposite. Due to the efficiency and maintainability reasons discussed in section~\ref{mem_const}, we adopt the static approach. To accommodate sacrificing the exact details of the memory management optimizations and allocation details, such as fragmentation and temporary memory for local variables, we spare 10\% of the device memory and constrain ourselves to the remaining 90\%. This threshold was sufficient to successfully run all our experiments without going out of memory (OOM). Nevertheless, this ratio might need to be tuned and it is the only parameter of \algo{} that needs tuning. 
%Moreover, an independent cost model can be developed to get an accurate estimate of this threshold and pass this value to \algo{}.

As shown in Figure~\ref{fig:imp}, %the collected data is passed to our algorithm, and 
the output of our algorithm is a single file containing model operations placement as key-value pairs. Each key is an operation-node name and the value is the device on which the operation should be allocated. To control the placement at operation-node granularity, %a small modification is done in its
the TensorFlow back-end reads the node-to-device assignment from the placement file generated by our algorithm. 

{\bf \algo{} on multiple nodes:}
%\didem{content here is not really assumptions or restrictions. Should we name the section as hybrid parallelism or Relations to Data Parallelism? or Extending ParDNN to multiple nodes?}
%\wahib{Assumptions and limitations on the type of models we work with}
 Despite the capability of designing \algo{} to partition a DNN on multiple nodes, in this work we assume a single node where the processing elements are identical GPUs connected to a common host. This is because  the number of GPUs per node has been steadily increasing over time. For instance, 
 %the three GPU-accelerated next generation US DoE machines include $8$ or more GPUs per node. In addition, 
 systems with  $16$ or more GPUs per node are in production (e.g. NVIDIA DGX SuperPOD). 
 %$20^{th}$ fastest system in November $2019$ Top500 list). 
As suggested %for further scaling 
by many state-of-the-art works~\cite{shoeybi2019megatron,Rajbhandari2019ZeROMO,huang2018gpipe},
 we argue that a hybrid approach of data parallelism across compute nodes 
 %(i.e. samples split between nodes) 
 and using \algo{} inside the compute node is a practical choice. This approach benefits from the efficiency and non-invasiveness of our method in tackling the memory capacity issue at the node-level, while also harnessing the weak scaling properties of data parallelism across the nodes.
 %Recent advancements in other approaches, namely 

\section{Results}
\label{sec:results}

\edit{
This section is organized into two parts. First part compares the performance of \algo{} against related work: critical path based heuristics, explicit model parallelism, redundant recompute, and an out-of-core method. The second part evaluates the scaling of \algo{}. % in terms of performance and overhead. 
Key findings of each part are as follows: 
 \paragraph{\bf Comparison with Related Work} (i)~\algo{} outperforms other critical path-based heuristics for computational graph placement ~\cite{kim1988general, mayer2017tensorflow}. Moreover, its empirical overhead is much lower than alternative well-performing heuristics. Replacing Step-1 of~\algo{} with other heuristics results in a significant performance drop or huge increase in the overhead. (ii)~\algo{} outperforms the distributed tensor computation framework, Mesh TensorFlow \cite{Mesh-TF} and provides much higher user productivity. (iii) \algo{} %in small batch sizes 
    outperforms Gradient Checkpointing~\cite{Yaroslav_Bulatov_2018} combined with data parallelism in many cases, yielding up to 2.7x speedup. More importantly, \algo{} enables training models where applying Gradient Checkpointing result in out of memory (OOM)
    even with a batch size of 1. (iv) \algo{} outperforms CUDA Unified Memory for all configurations and GPU counts. %numbers of GPUs.
    
 \paragraph{\bf Scaling} (i) For the same number of GPUs, \algo{} enables the use of more than 9x batch size on average over the maximum possible with data parallelism. %, for the tested five models. 
    (ii) Superlinear speedup in most models and configurations is observed  going from one GPU to $16$ GPUs.
} 

\edit{Using either of the memory heuristics gives similar performance (running time). Hence, we only report the results with the memory Heuristic-II for the sake of brevity.}

%\fareed{(ii) Due to the efficient partitioning of \algo{}, surprisingly, \algo{} provides speedup of up to 3.4x over DP for the same batch size and number of GPUs. This not only makes \algo{} a solution for models bounded by memory capacity, but also a viable alternative to DP.}

\subsection{Environment, Models, and Datasets}
We conducted all our experiments on a NVIDIA DGX-2 with $16$ Tesla V100 SXM3 32GB GPUs connected via NVSwitch. % with 300 GB/sec bandwidth between them. 
The throughput measurements are conducted over the interval between the $100^{th}$ and the $150^{th}$ training iterations to get stable results.
We use TensorFlow 1.15 and CUDA 10.0.

%Transformer = Transformer-4096-1024-64
\begin{table}
\begin{center}
\scriptsize{
\caption{Specifications of Models Datasets. (C)HSD: (Character) Hidden State Dimension, SL: Sequence Length, ED: Embedding Dimensions, RU: Residual Units, WF: Widening Factor, MD: Model Dimension, FS:Filter Size, P\_SZ: patch size.}
\label{tab:models}
\setlength\tabcolsep{3pt} % default value: 6pt
\resizebox{\linewidth}{!}{
\begin{tabularx}{\linewidth}{|c|c|c|c|c|c|c|c|}
%\hline
\cline{1-7}
\multirow{2}{*}{Model / Dataset} & \multirow{2}{*}{Acronym} & \multirow{2}{*}{\#Layers} & \multirow{2}{*}{HSD} & \multirow{2}{*}{SL} & \#Para.&\#Graph%&\multirow{2}{*}{Dataset}
\\
&&&&&($10^9$)&Nodes\\
\cline{1-7}
\multirow{3}{*}{\shortstack{RNN for Word-Level \\ Language~\cite{word-rnn-tensorflow} / \\ Tiny  Shakespeare~\cite{tinyshakespeare_2015}}}
                  &Word-RNN   
                  &8
                  &2048
                  &28
                  &0.34
                  &10578
%                  &\multirow{2}{*}{Tiny  Shakespeare~\cite{tinyshakespeare_2015}}
                \\%\cline{2-7}
                  &Word-RNN-2
                  &\edit{32}
                  &\edit{2048}
                  &25
                  &\edit{1.18}
                  &\edit{39074}
%                  &
                     \\%\cline{2-7}
                   & & & & & & 
                  \\\cline{1-7}
 &  & & CHSD & ED &  & \\
\cline{1-7}
\multirow{3}{*}{\shortstack{Character-Aware Neural\\Language Models~\cite{kim2016character} / \\Penn Treebank (PTB)~\cite{marcus1994penn}}}
                  &Char-CRN   
                  &8
                  &2048
                  &15
                  &0.23
                  &22748
               %&\multirow{2}{*}{Penn Treebank (PTB)~\cite{marcus1994penn}}
                  \\%\cline{2-7}
                  &Char-CRN-2
                  &32
                  &2048
                  &15
                  &1.09
                  &86663
                   \\%\cline{2-7}
                   & & & & & & 
%                  &
                  \\\cline{1-7}
 &  & &\#RU&  WF  & & \\
\cline{1-7}
\multirow{2}{*}{\shortstack{Wide Residual Net.~\cite{zagoruyko2016wide} / \\CIFAR100~\cite{krizhevsky2009learning}}}
                  &WRN 
                  &610
                  &101  
                  &14
                  & 1.91
                  & 187742
                   %&\multirow{2}{*}{CIFAR100~\cite{krizhevsky2009learning}}
                  \\%\cline{2-7}
                  &WRN-2
                  &304
                  &50
                  &28 
                  & 3.77
                  & 79742
                  \\\cline{1-7}
%                  &\edit{WRN-3}
%                  &\edit{610}
%                  &\edit{101}
%                  &\edit{6}
%                  & \edit{0.35}
%                  & \edit{187742}
%                  &\edit{ImageNet~\cite{imagenet_cvpr09}}
%                  \\\cline{1-8}
 &  & & HSD & MD & & \\
\cline{1-7}
\multirow{2}{*}{\shortstack{Transformer~\cite{vaswani2017attention} / IWSLT'16 \\ German–English corpus~\cite{cettolo2016iwslt}}}
                  &TRN   
                  &\edit{52}
                  &\edit{4098}
%                  &32K
                  &\edit{2048}
%                  &16
                  &\edit{1.99} 
                  &\edit{204792}
                  
% &\multirow{2}{*}{\shortstack{IWSLT 2016 \\ German–English corpus~\cite{cettolo2016iwslt}}} 
                  \\%\cline{2-7}
                  &TRN-2
                  &48 
                  &8192 
%                   &32K
                  &2048
%                  &16
                  &5.1
                  &160518
%                  &
                  \\\cline{1-7}
 &  & HSD &  FS & P_SZ && \\
\cline{1-7}
\multirow{2}{*}{\shortstack{Eidetic 3D LSTM\cite{wang2018eidetic} / \\ Moving MNIST digits~\cite{srivastava2015unsupervised}}}
                  &E3D 
                  &320  
                  &5 
                  &4
                  & 0.95
                  & 55756 
                  %&\multirow{2}{*}{}
                  \\%\cline{2-7}
                  &E3D-2
                  &512
                  &5 
                  &8
                  & 2.4
                  & 55756 
%                  &
                  \\\cline{1-7}
\end{tabularx}}
%	\vspace{-1em}
	}
	\end{center}
\end{table}

\subsection{Models and Datasets}
To demonstrate our results we experimented with five large models representing four main tracks of DL applications: image classification, translation, video prediction, and language modeling. All models  and datasets used in the experiments are listed in Table~\ref{tab:models}. We focus our analysis on the performance of \algo{}, rather than pursuing the accuracy as \algo{}
%. It is important to mention that \algo{} operates at model execution level and 
has no effect on the learning aspect of the model. More specifically, the convergence and accuracy are mainly affected by changing the batch size and other hyper-parameters, and our algorithm does not alter the model and its hyper-parameters in any fashion. \algo{} changes the placements of the operations on devices after the computational graph has already been generated by the framework. 

%From the language modeling, 
We use \textbf{Word-RNN} a multi-layer Recurrent Neural Network for word-level language inspired by the character-level modeling~\cite{sutskever2011generating}.  Character-Aware Neural Language Models (\textbf{Char-CRN})~\cite{kim2016character}.
Both models can be enlarged by increasing the number of layers or the hidden state size.
\textbf{WRN}~\cite{zagoruyko2016wide} is a widened version of the residual network model. In WRN the number of residual units and the width of the convolutional layers can be configured. The model size grows quadratically when widened. WRN has been achieving better accuracy when the model is widened~\cite{zagoruyko2016wide}. \textbf{TRN} (Transformer)~\cite{vaswani2017attention} can be enlarged by increasing the number of layers, which deepens the model, and by widening the inner-layer dimensionality. Deeper~\cite{huang2019gpipe} and wider~\cite{vaswani2017attention} configurations of Transformer are shown to give higher accuracy.
\textbf{E3D} is Eidetic 3D LSTM~\cite{wang2018eidetic} for video prediction. 
{\em E3D} can be enlarged by increasing the number of the hidden state channels on the memory dimensions.

We experimented with models under two main use-cases of \algo{}. First, model instances that {\em fit into a single device memory only when very small batch size is used}. Small here is relative to the numbers used by the DL community and reported in the literature. In such a case, \algo{} provides a qualitative advantage over data parallelism (DP), which splits the input over different GPUs that hold the replicas of the model. 
The second use-case is {\em model instances that do not fit into a single GPU memory} even with small batch sizes.  These are the larger variants of each model, as shown in Table \ref{tab:models} (e.g., Word-RNN-2).~\edit{It is important to emphasize that the motivation of \algo{} is on cases when the model can not fit in memory or fits with very small batch size, which has become a major challenge for large models. Hence the experiments focus on weak scaling,  not on strong scaling. If the model fits in memory with large enough batch size, then we would not suggest using graph-based methods.}

\subsection{Comparison with Related Work} %Methods for Breaking the Memory Limit}
We compare \algo{} with two graph (critical path)-based methods and three different state-of-the-art approaches used to circumvent the memory limitation when training DNNs. 
We compare with (a) Linear Clustering~\cite{kim1988general} and Critical Path~\cite{mayer2017tensorflow} methods, (b) Mesh-TensorFlow~\cite{Mesh-TF} for explicit model parallelism,  (c) Gradient Checkpointing ~\cite{Yaroslav_Bulatov_2018} in combination with data parallelism for redundant recompute and (d) CUDA Unified Memory for out-of-core computing. 
%For explicit model parallelism, we compare with Mesh-TensorFlow~\cite{Mesh-TF} that extends TensorFlow. For redundant recompute, we compare with gradient checkpointing~\cite{Yaroslav_Bulatov_2018} in combination with data parallelism. For out-of-core computing, we compare with CUDA Unified Memory. 
Although there exists other graph-based solutions, we cannot directly compare with them either because we are not aware of any open source implementation ~\cite{mirhoseini2018hierarchical} or 
the implementation is available for MXNet only ~\cite{wang2019supporting} and cannot support all the operations used in Tensorflow. It is worth mentioning, however, that \algo{} takes no more than $2$ minutes for the largest configuration we tested, in comparison to $10$s of hours reported by the other graph-based methods~\cite{mirhoseini2018hierarchical}. In addition, \algo{} working on models 2.3x as large as what the these methods experimented with~\cite{mirhoseini2018hierarchical, 10.5555/3305890.3305932}. 
%\fareed{What does it mean  large as Xx the other methods or how can we tell what is Xx? Is it by dividing the number of parameters of a model over another?}\wahib{Yes. you could use the number of parameters or number of nodes in the graph.}  
%we do not compare to other graph-based solutions. 
%Regarding, we are not aware of any open source implementation.  For~\cite{wang2019supporting}, the implementation is only available for MxNet. \didem{Fareed?}
%it is inapplicable for Tensorflow. 
%\wahib{Note to self: continue the phrase in a convincing argument}.

 \begin{figure}[!t]
\centering
\includegraphics[scale=0.52]{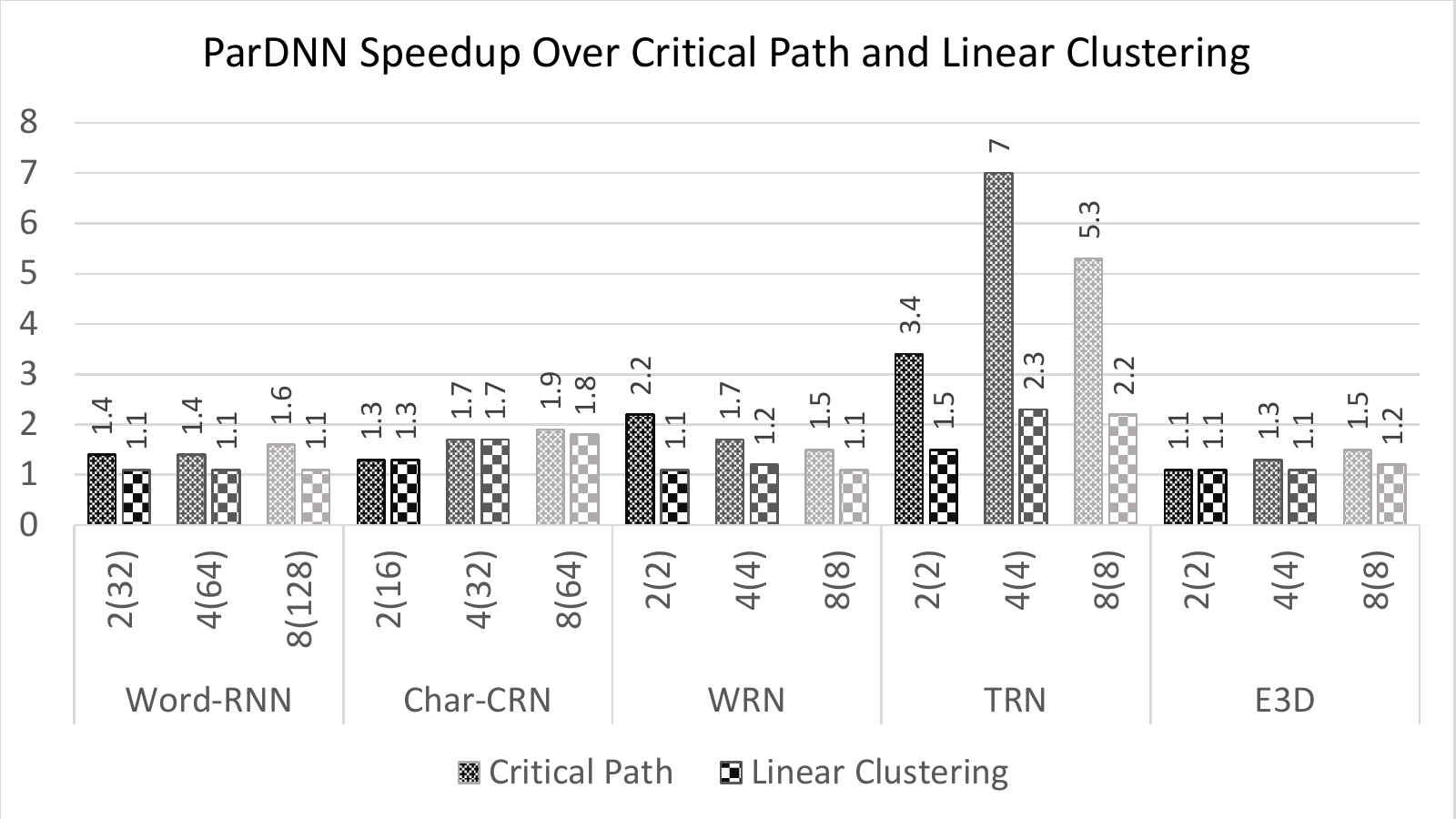}
\caption{\edit{\algo{} speedup over Linear Clustering~\cite{kim1988general} and Critical -Path (CP) heuristic~\cite{mayer2017tensorflow}. x-axis: Number of GPUs (Batch Size)}}
\label{fig:cp_lc}
 \end{figure}
 
\subsubsection{\edit{Graph-based methods: Linear Clustering and CP}}
\label{sec:lc_cp}
\edit{
In~\cite{mayer2017tensorflow}, the authors proposed a set of heuristics to partition Tensorflow graphs among multiple devices. Among the proposed heuristics, a critical path-based heuristic, referred as CP, achieves the best results in all of their experiments. Authors performed event-based simulations to evaluate their partitioning. We applied their heuristic on our graphs, extracted the partitioning before the simulation step and fed them to Tensorflow. As Figure~\ref{fig:cp_lc} shows, \algo{} outperforms CP using all models on
all device counts.

To demonstrate  that our choice of using a multi-staged approach over a high-complexity single heuristic  does not harm the quality of the partitioning,
we compare \algo{} with Linear Clustering (LC). To do a fair comparison, we implemented LC with GLB and Earliest Estimated Time First 
(EST First)~\cite{wang2018list} as a task ordering heuristic since this combination gave the best results. We post-processed the result to meet the Tensorflow placement constrains. \algo{} outperforms LC in all experiments, even though it sacrifices nodes' weighted-level recalculation after the $K$'th iteration, thanks to the novel mapping heuristic. Moreover, while it took \algo{} $\sim 2$ minutes to produce the placement for the largest graph (TRN with 205K nodes), it took linear clustering $\sim 12$ hours. CP, on the other hand, is quite fast. It partitions all the graphs within seconds. Both \algo{} and LC are far better than CP, this is because CP assigns all the nodes outside of the critical path to the least loaded devices without further grouping/clustering. In other words, it does not consider locality outside the critical path.
}

\begin{figure}[!t]
\centering
 \includegraphics[scale=0.68]{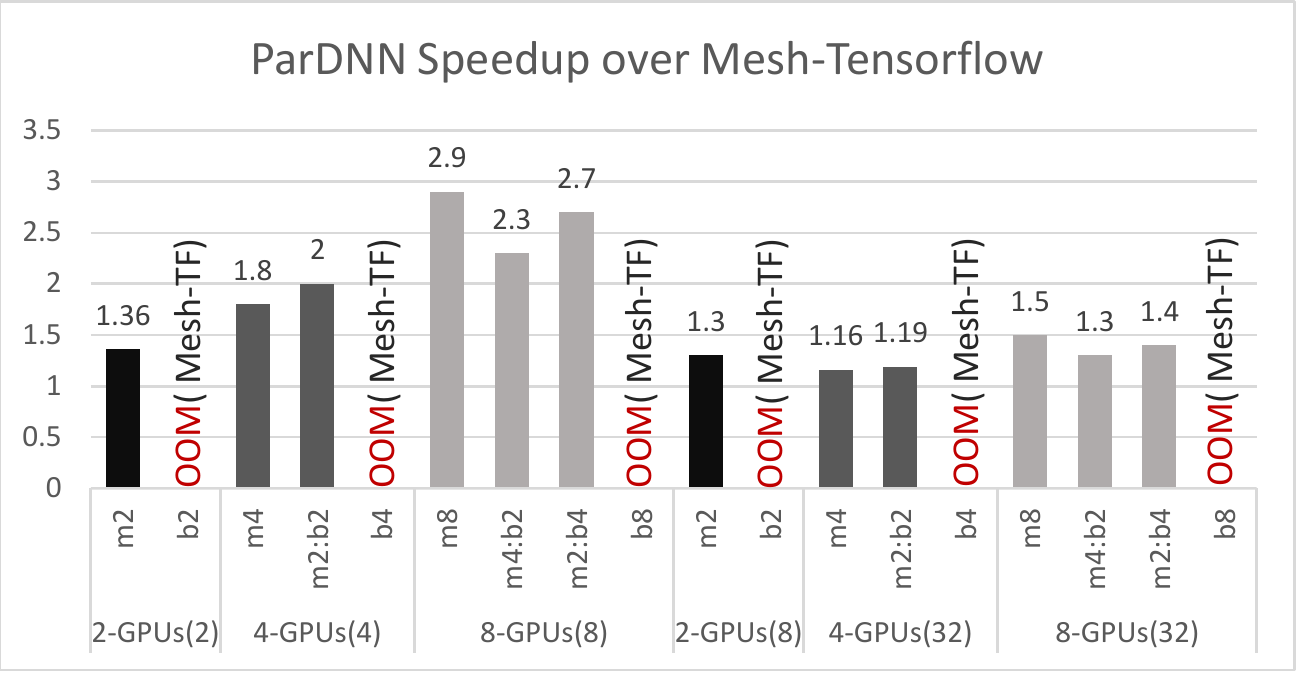}
\caption{\edit{\algo{} speedup over Mesh Tensorflow. X-axis: number of GPUs (batch size), and the possible permutations permitted by Mesh-Tensorflow.}
}
 \label{fig:mtf}
 \end{figure}

\begin{figure*}[!t]
\centering
 \includegraphics[width=\textwidth]{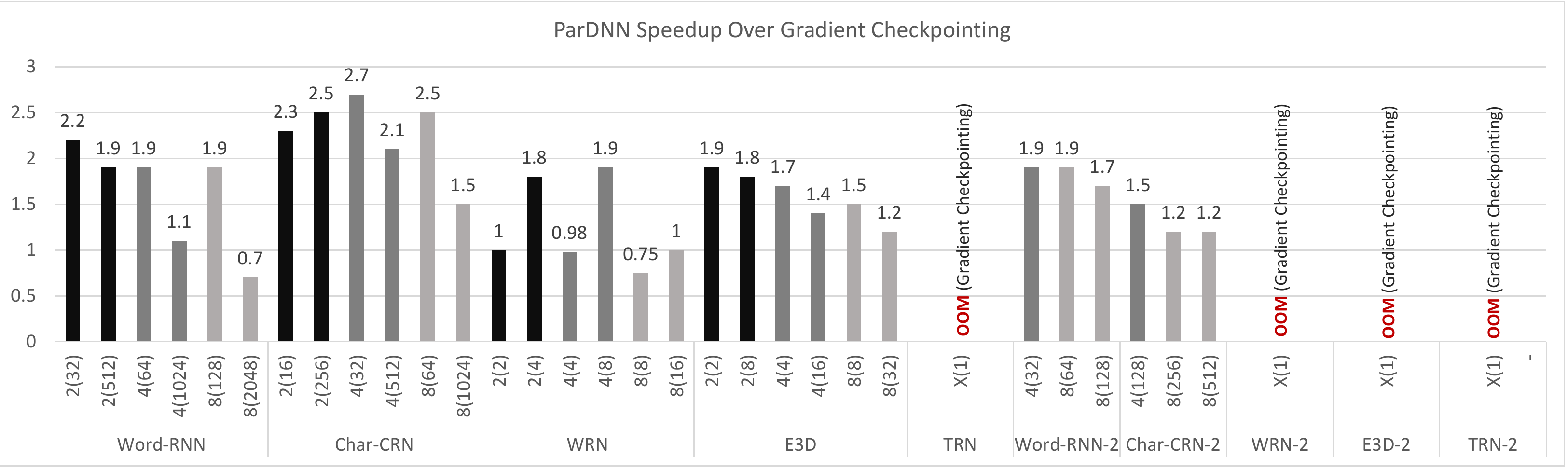}
\caption{\edit{\algo{} speedup over gradient checkpointing combined with data parallelism to run on multiple GPUs (X-axis: Number of GPUs (Batch Size)). Two batch size scalings were used (a) doubling the batch size with the number of devices, (b) if both techniques allow larger batches, the comparison is held using the maximum commonly trainable batch size. X indicates gradient checkpointing (combined with data parallelism) could not produce a valid solution regardless of GPU count.} 
}
 \label{fig:mtf_gcp}
 \end{figure*}
 
\subsubsection{\edit{Mesh-TensorFlow}}
Mesh-TensorFlow~\cite{Mesh-TF}, an extension to TensorFlow, was proposed to overcome the memory limitations of a single device. It permits specifying a general class of distributed tensor computations. We compare the performance of \algo{} with Mesh-TensorFlow using the Transformer model which the original authors used to demonstrate the scaling~\cite{Mesh-TF}.   
Figure \ref{fig:mtf} shows the speedup of \algo{} over Mesh-TensorFlow using $2$, $4$ and $8$ GPUs. We report all permutations~\cite{tensor2tensor} \edit{with both regular weak scaling and the maximum trainable batch size for most of these permutations}.
\edit{\algo{} outperforms Mesh-TensorFlow in both cases. The performance gap is much larger with smaller batch sizes. This is because when the batch size is small, GPUs are underutilized, and Mesh-Tensorflow splitting across model dimension, batch dimension or both creates even smaller kernels exacerbating underutilization. 
However, when the batch sizes are large enough, the benefit of the parallelism created by Mesh-Tensorflow outweighs a possible minor underutilization leading to a relatively good performance.} Moreover, unlike Mesh-TensorFlow (a) \algo{} requires no knowledge about the DNN structure by the user, while with Mesh-TensorFlow it is the responsibility of the user to rewrite the model using Mesh-TensorFlow  syntax. (b) \algo{} entirely automates the partitioning, while with Mesh-TensorFlow users have to manually specify the tensor-dimensions to be split across a multi-dimensional processor mesh and finding the best assignment is an NP-hard problem. (c) Mesh-TensorFlow has a non-negligible pre-run overhead which doubles when doubling the number of GPUs reaching $ \sim 1$ hour for $8$ GPU assignment.

\subsubsection{Redundant Recompute: Gradient Checkpointing}
%\didem{this part needs to be rewritten based on the new results}
Gradient checkpointing~\cite{Chen2016TrainingDN} 
%is a general approach that works with both convolutional and recurrent neural networks. It 
enables DNN training  with a sublinear memory cost (${O(\sqrt{N})}$) when training an ${N}$ layer network by recomputing the activations during backpropagation, instead of holding the forward pass results. In our comparison, we use a TensorFlow-based open-source implementation~\cite{Yaroslav_Bulatov_2018}. 
Figure~\ref{fig:mtf_gcp} shows the speedup of \algo{} over gradient checkpointing when combined with data parallelism to run on multiple GPUs.~\edit{For \algo{} and checkpointing, we used both regular weak scaling and the common largest possible batch sizes}. \algo{} outperforms gradient checkpointing in most cases. In few cases, checkpointing is better than \algo;
% Checkpointing combined with DP outperforms \algo{} in some cases; 
this happens mainly when the degree of parallelism inherent in the graph is not sufficient to  fully utilize all the GPUs when the model is partitioned, or when checkpointing enables pushing large enough batch-size that guarantees satisfactory utilization of these GPUs. 
However, more importantly, \algo{} is qualitatively superior to gradient checkpointing since it enables the training of models \edit{by using multiple GPUs where checkpointing fails to make them fit in device memory, even when using a batch size of one}. For example, Figure~\ref{fig:mtf_gcp}(b) shows several configurations where gradient checkpointing goes out-of-memory at the batch size of one. Moreover, the overhead of gradient checkpointing can be up to $5$ hours~\cite{Yaroslav_Bulatov_2018}.
%GCP recomputes the intermediate results during backpropagation instead of keeping the forward pass results around.}
%drops some of the intermediate results, and during the backpropagation, the dropped results are re-computed by running a forward pass from the closest recorded results.  }

 \begin{figure}[!t]
\centering
\includegraphics[scale=0.7]{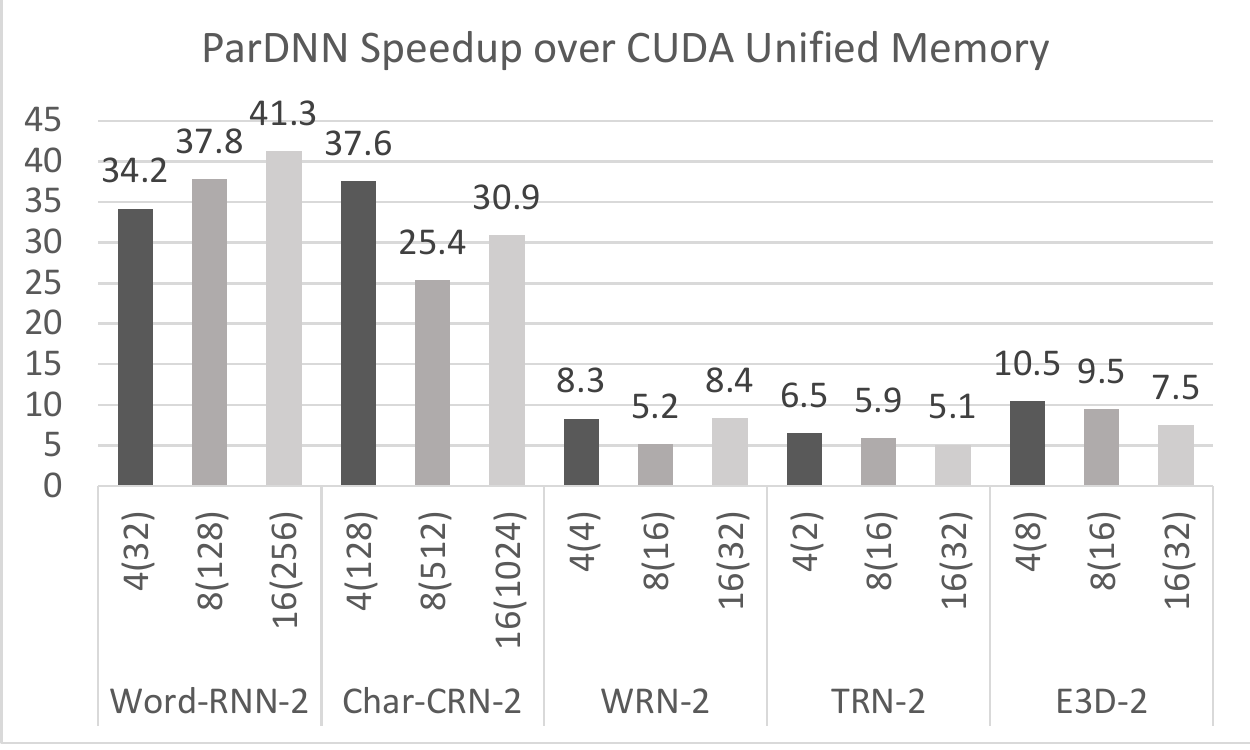}
\caption{\edit{\algo{} speedup over CUDA Unified Memory (UM) using the large models. X-axis: Number of GPUs (Batch Size)}}
\label{fig:um}
 \end{figure}
 
\subsubsection{Out-of-core: CUDA Unified Memory}
Figure~\ref{fig:um} shows the speedup of \algo{} over CUDA Unified Memory (UM). UM, to the authors knowledge, is the only out-of-core solution that has an available Tensorflow implementation.~\algo {} outperforms UM in all cases. Although UM allows pushing large batch sizes what enhances GPU utilization, its performance degrades in many cases when increasing the batch size due to the larger device-host communication cost and the page faulting penalty~\cite{awan2018oc}.

%\subsection{Models Fitting into a Single GPU}
\subsection{Scaling Studies}
%We experimented with models under two main use-cases of \algo{}. First, model instances that \emp{ fit into a single device memory only with very small batch sizes}. Small here is relative to the numbers used by the DL community and reported in the literature. In such a case, \algo{} provides a qualitative advantage over data parallelism (DP), which splits the input over different GPUs that hold the replicas of the model. 
%The second use-case is {\em model instances that do not fit into a single GPU memory} even with small batch sizes.  These are the larger variants of each model, as shown in Table \ref{tab:models}.~\edit{We would like to emphasize that our motivation and focus is on cases when the model can not fit in memory or fits with very small batch size, which is becoming a major challenge for large models. Hence in this paper, we focus on weak scaling and not strong scaling. If the model fits in memory with large enough batch size to do strong scaling, then we would not suggest using graph-based methods.}
%with up to $5.1B$ parameters in {\em TRN-2}.
%having from $1.09B$ in {\em Char-CRN-2} to $5.1B$ parameters in {\em TRN-2}.

\begin{table}
\footnotesize{
\caption{\edit{Maximum batch sizes (bsz) made possible by \algo{}. Bsz on a single GPU is the maximum that could fit without triggering OOM. 
Table also shows the multiplier by which \algo{} could increase the bsz over ideal data parallelism (DP). For use-cases-1, DP is assumed to applied on top of a single GPU reference point. 
For use-cases-2, \algo{} enables $\geq$ 4-GPU assignment and DP is assumed to be applied on top of 4-GPU reference point. We report the values enabled by both of the memory heuristics.}}}
\renewcommand{\tabcolsep}{3pt}
\resizebox{\linewidth}{!}{
\begin{tabularx}{\linewidth}{|c|c|c|c|c|c||c|c|c|c|c|}
\cline{1-11}
 & \multicolumn{5}{c||}{\footnotesize{Batch Size Scaling}} & \multicolumn{5}{c|}{\footnotesize{Increase Over Ideal DP}} \\\cline{1-11}
\footnotesize{Model / \#GPUs}	&	1	&	2	&	4	&	8	&	16	&	1	&	2	&	4	&	8	&	16	\\ \cline{1-11}
\footnotesize{Word-RNN}	&	16	&	512	&	1024	&	2048	&	2048	&	1x	&	16x	&	16x	&	16x	&	8x	\\
\footnotesize{Char-CRN}	&	8	&	256	&	512	&	1024	&	2048	&	1x	&	16x	&	16x	&	16x	&	16x	\\
\footnotesize{WRN}	&	1	&	4	&	16	&	\edit{16}	&	\edit{32}	&	1x	&	2x	&	4x	&	\edit{2x}	&	\edit{2x}	\\
\footnotesize{TRN}	&	1	&	\edit{8}	&	\edit{32}	&	\edit{64}	&	\edit{128}	&	1x	&	\edit{4x}	&	\edit{8x}	&	\edit{8x}	&	\edit{8x}	\\
\footnotesize{E3D}	&	1	&	\edit{8}	&	16	&	\edit{16}	&	\edit{32}	&	1x	&	\edit{4x}	&	4x	&	\edit{2x}	&	\edit{2x}	\\ \cline{1-11}
\footnotesize{Word-RNN-2}	&	--	&	--	&	32	&	128	&	256	&	--	&	--	&	1x	&	\edit{2x}	&	\edit{2x}	\\ 
\footnotesize{Char-CRN-2}	&	--	&	--	&	128	&	512	&	1024	&	--	&	--	&	1x	&	2x	&	2x	\\
\footnotesize{WRN-2}	&	--	&	--	&	4	&	16	&	32	&	--	&	--	&	1x	&	2x	&	2x	\\
\footnotesize{TRN-2}	&	--	&	--	&	\edit{2}	&	16	&	32	&	--	&	--	&	1x	&	4x	&	4x	\\
\footnotesize{E3D-2}	&	--	&	--	&	8	&	16	&	32	&	--	&	--	&	1x	&	1x	&	1x	\\
%\edit{WRN-3}	&	\edit{--}	&	\edit{--}	&	\edit{2}	&	\edit{4}	&	\edit{8}	&	\edit{--}	&	\edit{--}	&	\edit{1x}	&	\edit{1x}	&	\edit{1x}	\\
\cline{1-11}

\end{tabularx}
}
\label{tab:batch_scaling}
\end{table}

\begin{figure*}[!t]
\begin{center}
\includegraphics[width=\textwidth]{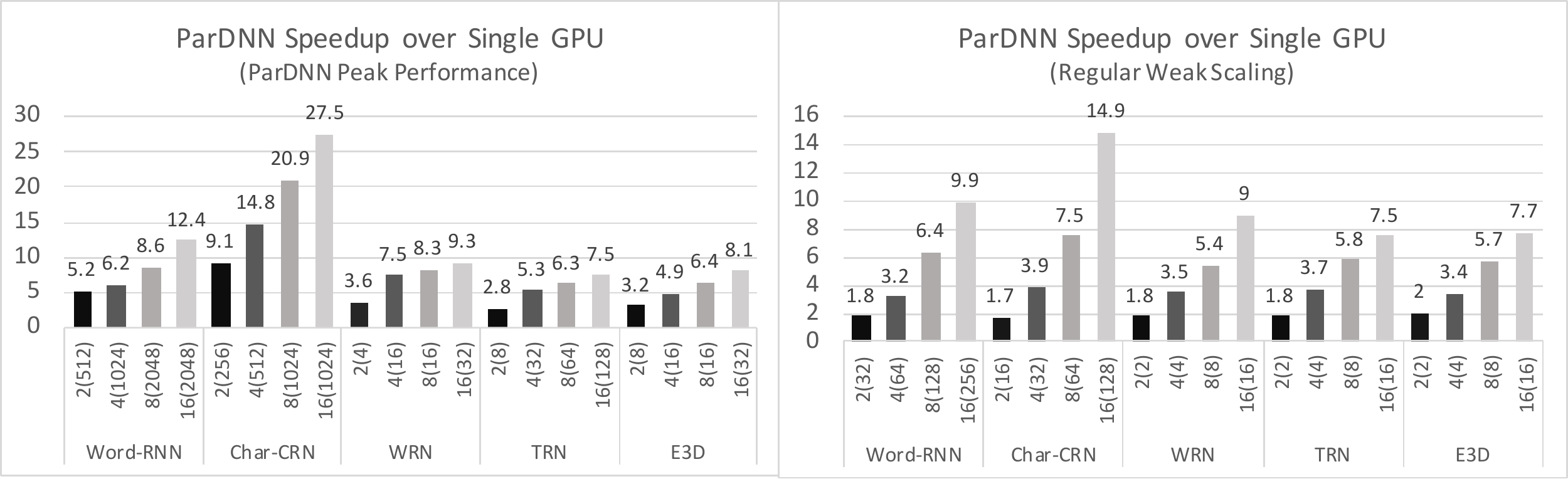}
\hspace{4cm} (a) \hspace{8cm} (b)
\caption{\edit{\algo{}  speedup  over  a  single  GPU  using 2, 4, 8 and 16 GPUs. X-axis: Number of GPUs (Batch Size)}}  
\label{fig:single}
\end{center}
\end{figure*}
 
\subsubsection{{\bf Batch Size Scaling}}
Training with large batch sizes offers more parallelism and drastically reduces the overall training  time. Authors 
in ~\cite{goyal2017accurate} proposed a method to scale  batch sizes, which reduced the
training of RESNET-50 on ImageNet to one hour. 
Another work harnessed very large batch sizes to reduce BERT training time from 3 days to 76 mins ~\cite{you2019large}.
\algo{} enables superlinear scaling of the batch sizes while increasing the number of GPUs. Table 4 shows the batch size scaling for all of our experiments.  %except one, {\em E3D-2}. 
We could increase the batch size by up to $256x$ for use-cases-1 and $16x$ for use-cases-2 going from one to 16 GPUs. This gives \algo{} a qualitative advantage even for models that fit into a single GPU 
since \algo{} enables training with much larger batch sizes than what can be achieved with DP. 

\algo{} achieves superlinear scaling of the batch size firstly because with \algo{}, the parameters are not replicated but distributed. A large fraction of the memory consumed by the large models is to store the parameters and variables that survive through iterations. For instance, for $1.91$ billion parameter {\em WRN}, TensorFlow allocates around 8GB for those variables. Using \algo{} these parameters are distributed, but with DP they need to be replicated. %, which wastes considerable amount of memory. 
Secondly, for some operations, the memory consumption does not scale linearly with the batch size. For example, in {\em Word-RNN} and {\em Char-CRN}, the outputs of matrix multiplication  operations have the largest memory consumption ratio. When doubling the batch size,  the memory consumption by matrix multiplication results increases by only $\sim25\%$. This is because the batch size might be the inner dimension for many of these multiplications, when multiplying a matrix of dimensions 
$a * ${\em batch\_size} by another of {\em batch\_size}$ * b$, the result has the dimensions of ${a * b}$ regardless of the {\em batch\_size}. So the memory allocated to store the output of that operation does not increase, and this effect propagates to its decedents that will take its output as their input.%\wahib{Well done Fareed! Good explanation of the gap (I just did minor edits)}
 
\begin{figure}[!t]
\begin{center}
\includegraphics[scale=0.5]{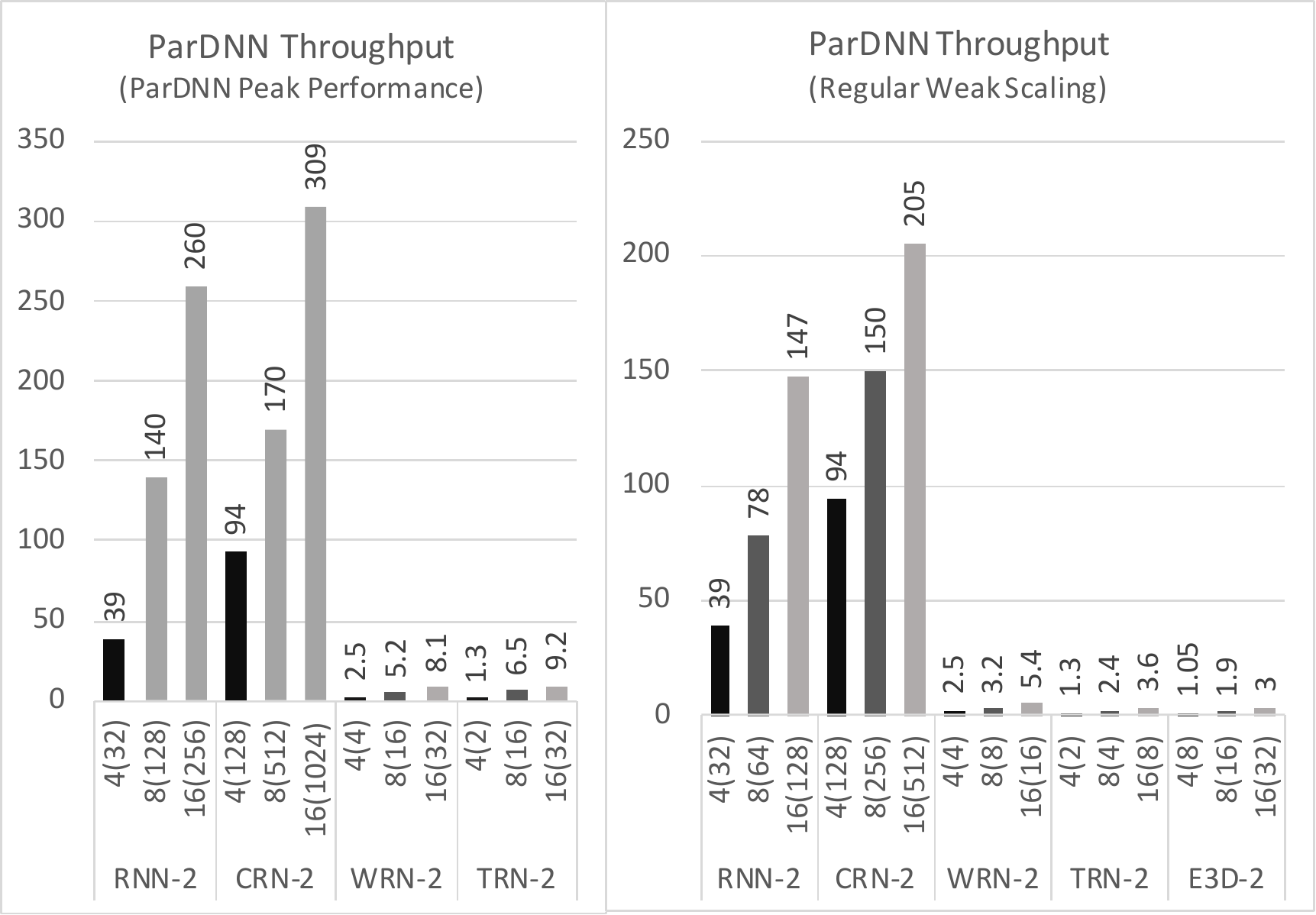}
\hspace{4cm} (a) \hspace{4cm} (b)
\caption{\edit{Throughput  and  scaling up to 16 GPUs with use-case-2 models. X-axis: Number of GPUs (Batch Size)}} %$2^{nd}$ digit after the decimal point. }  
\label{fig:throughput}
\end{center}
\end{figure}

\subsubsection{\bf GPU Count Scaling}
Figure \ref{fig:single} and \ref{fig:throughput} show the speedup over a single GPU for small models and  the throughput scaling of \algo{} for large models, respectively. \edit{We used two different sets of batch sizes to demonstrate scaling.  The first demonstrates scaling with the batch sizes at which \algo{} achieves the peak performance, the vast majority of these batch sizes are the ones listed in Table 4. The second doubles the batch size with the number of GPUs (regular weak scaling).} 

In Figure \ref{fig:single} (a), \algo{} shows a substantial improvement on $2$ GPUs and superlinear speedups up to $4$ GPUs for all the models. The sharp performance increase  happens because, in addition to the parallelism introduced by adding more GPUs, pushing larger batches while doubling the number of GPUs, improves the device utilization considerably.

\begin{figure}[!t]
\begin{center}
\includegraphics[scale=0.60]{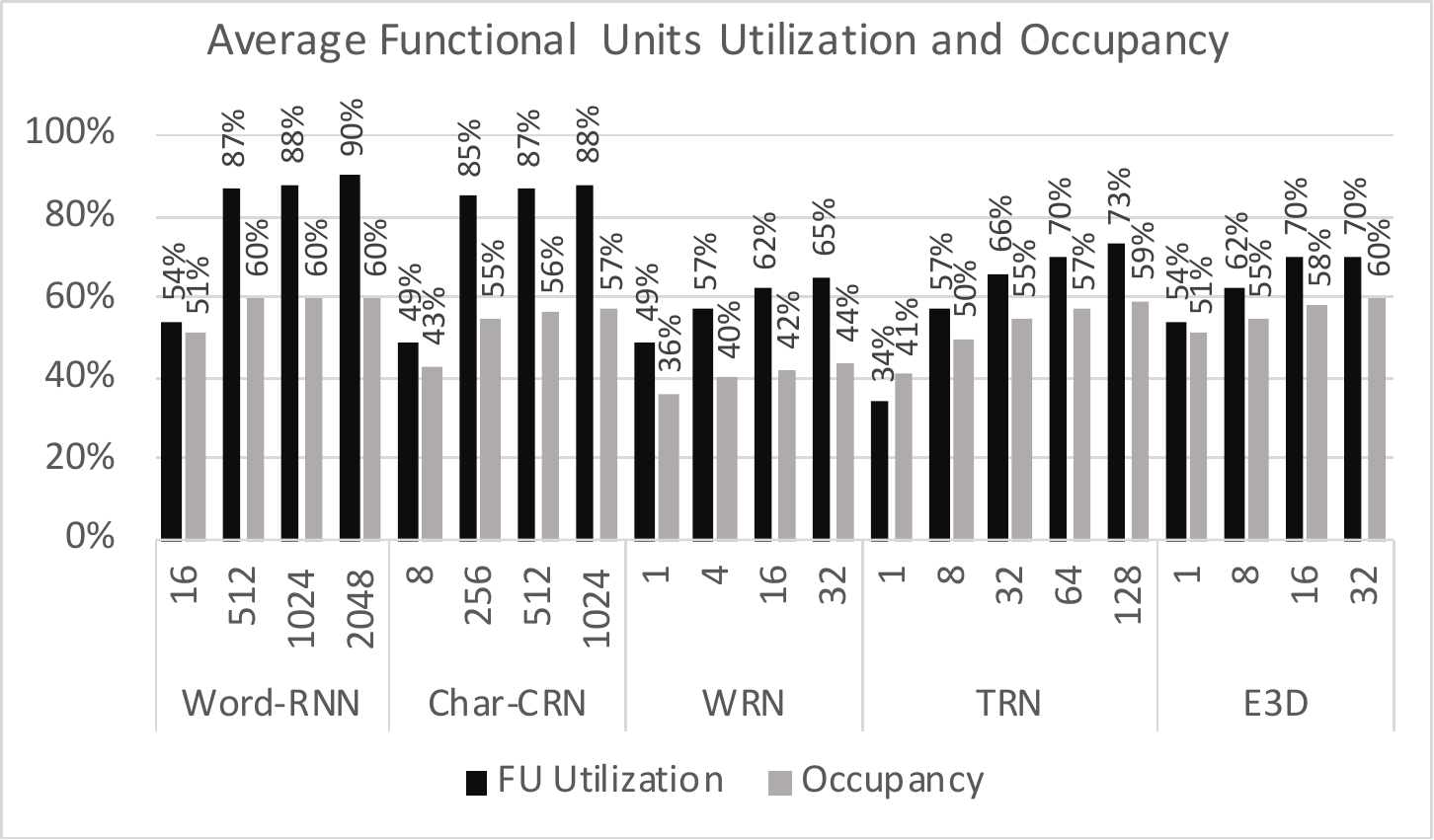}
\caption{\edit{Average functional unit utilization and GPU occupancy with different batch sizes. X-axis: Batch Size}} %$2^{nd}$ digit after the decimal point. }  
\label{fig:fu_occ}
\end{center}
\end{figure}

\edit{ Figure~\ref{fig:fu_occ} shows the improvement in GPU functional unit utilization and GPU occupancy measured by using {\em nvprof} by Nvidia. We observe a similar leap where the values of these metrics increase considerably with the batch sizes used on 2, and in some cases, 4 GPUs. With the batch sizes used on $8, 16$ GPUs, the utilization improves at a very small rate. Hence the scaling depends more on the inherent DoP (average degree of parallelism) in the graph and CCR (Table \ref{tab:dop_ccr}). More specifically, the models with higher DoPs scales better since introducing more GPUs exploits the parallelism. However, models with low DoP would result in small improvement in utilization since there is no parallelism inherent in the graph to be harnessed by introducing more GPUs. 

{\em Char-CRN} has a large DoP, hence it continues to give superlinear speedups up to $16$ GPUs. However, its scaling is not perfect due to having very high CCR, which results in having more communication links between the partitions with higher GPU counts. {\em Word-RNN} scales reasonably between $4$ and $16$ GPUs, thanks to its high DoP. However, the scaling is not perfect again due to the high CCR. {\em E3d} scales reasonably due to its medium CCR: the scaling is not ideal due to the relatively low DoP. Moreover, E3D experiences the lowest improvement in GPU utilization since its main operation is 3D convolution, which utilizes the GPU well enough ( $>50\%$ functional unit utilization) even with a batch size of $1$. Both {\em TRN} and {\em WRN} have high CCR and low DoP, hence they have the lowest improvement rate going from $4$ to $16$ GPUs. In Figure~\ref{fig:single} (b), the magnitudes of the achieved speedups are less due to smaller batch sizes which means less utilization (as a magnitude) compared to their counterparts in Figure~\ref{fig:single} (a). However, the scaling is better (going from X to 2X GPUs) up to $16$ GPUs. This is because most of the used batch sizes belong to the ranges at which the utilization considerably improves with larger batches (before the plateau in Figure~\ref{fig:fu_occ}).}

\edit{In Figure~\ref{fig:throughput} (a), going from $4$ to $8$ GPUs enables much larger batches in all cases. This in turn enhances the resource utilization and results in substantial throughput improvements. {\em Char-CRN-2} scales linearly up to $16$ GPUs due to its high DoP. The same applies for {\em Word-RNN-2}. {\em WRN-2} and {\ TRN-2} scale modestly from $8$ to $16$ due to the low DoP. But the scaling is better than in Figure~\ref{fig:single} since the batch sizes trainable with these models are located in the region at which the GPU utilization still improves with larger batch sizes (before the plateau). In Figure~\ref{fig:throughput} (b)~\algo{} enables linear scaling of the batch size for {\em E3D-2} with a performance scaling behavior similar to {\em E3D} for the same reasons.}

\begin{table}[t]
\begin{center}
\footnotesize{
\caption{\edit{Degree of Parallelism (DoP) and Communication to Computation Ratio (CCR) of the models.}}
\label{tab:dop_ccr}
\begin{tabularx}{0.5\linewidth}{|l|X|X|}
\hline
Model & DoP & CCR\\\hline
 Word-RNN & 10.45 & 14.5 \\\hline
 Char-CRN & 49 & 57\\\hline
 WRN & 1.16 & 13.02\\\hline
 TRN & 2 & 13.7\\\hline
 E3d & 3.3 & 1.12\\\hline
\end{tabularx}
}
\end{center}
\end{table}

\subsection{Overhead of \algo}
\algo{} has a negligible overhead thanks to the low complexity of each step.
%Step-1 (${O(K(\vert V \vert + \vert E \vert))}$). 
%Table~\ref{tab:overhead} shows the time for the longest partitioning instances among the different combinations of batch sizes, GPUs and, model configurations. 
The longest partitioning time among all the combinations of batch sizes, GPUs and model configurations used in this work was $2$ minutes in the case of partitioning {\em TRN-2} over 16 GPUs. The minimum time of $18$ secs was taken to partition {\em Word-RNN} over 2 GPUs.  Even though handling the memory overflow in case of memory heuristic-I 
%is used as heuristic-II has a low time complexity,} 
takes most of the overall partitioning time, 
%The fast partitioning time of \algo{} demonstrates that 
the time taken to handle memory overflow is much lower than the theoretical upper bound. This is because the complexity analysis of Step-2 of \algo{} depends on how many nodes need to be moved between clusters to address the overflow, which is much less than $\vert V \vert$ in practice. 
The average ratio of the  nodes moved in all our experiments is ${8\%}$.

%\algo{} scales favorably in comparison to other state-of-the-art static scheduling algorithms which have quadratic complexities. For example ETF of ${O(K\vert V \vert^2)}$ complexity~\cite{hwang1989scheduling} took $104$ seconds ($\sim2$ minutes) to schedule the small {\em WRN} with $11,744$ nodes, the time escalated to $4,072$ seconds ($> 1$ hour) with {\em TRN-2} of $160,519$ nodes. This trend suggests that when the models grow to have millions of nodes it will take hours if not a whole day to obtain a schedule.

\section{Related Work}
We summarize related work that touches on different aspects of \algo{}: from techniques that handle, or can alleviate, memory bottlenecks either using a single device by partitioning a DNN across multiple devices, to graph partitioning and scheduling.

{\bf Systems-level approaches}: 
Mirhoseini et al. proposed a reinforcement learning-based method to place dataflow graphs on multiple devices~\cite{10.5555/3305890.3305932, mirhoseini2018hierarchical}. This approach suffers from significant time and resource consumption. The proposed policy was trained for hours using $16$ workers to produce placements for models having less than $100K$ operations. A more efficient approach was proposed by Wang et al. in~\cite{wang2019supporting}. However, it requires a description language to specify computations and cannot describe all the operations used in DL. Moreover,
%(b) it depends on a limited partition and reduce pattern that requires each worker to perform a coarse-grained task identical to the original computation,
it partitions all operators and tensors across all workers, resulting in poor resource utilization. ~\edit{In~\cite{mayer2017tensorflow}, authors propose a set of practical heuristics to partition Tensorflow graphs. They concluded that critical path-based approaches yield the best performance.}

{\bf DL-level approaches}: 
Explicit model parallelism, where each worker is responsible for a subset of the layers, suffers from two major limitations: requiring complex cost models on case-by-case bases and leaving the partitioning burden to the programmer~\cite{narayanan2019pipedream}. Pipeline parallelism provides good resource utilization yet some implementations requires a single layer to fit in a single device~\cite{huang2019gpipe}, which may not be the case for models with 3D inputs~\cite{Mathuriya:2018:CUD:3291656.3291743}. While in others, extra memory overhead proportional to the size of the model weights is necessary to address the statistical efficiency issue, i.e. %and if not , it can delay and 
preventing model convergence~\cite{narayanan2019pipedream}.  %In~\cite{jia2018exploring, krizhevsky2014one, Naoya:sc19,shoeybi2019megatron} non-generic techniques were proposed to parallelize specific types of DL models that are  mainly used in the computer vision field.
In~\cite{jia2018exploring, krizhevsky2014one, Naoya:sc19,shoeybi2019megatron} non-generic techniques were proposed to parallelize specific types of DL models, some focusing on CNNs %~\cite{jia2018exploring, krizhevsky2014one, Naoya:sc19} 
while others relying on Transformer in their optimizations.  %~\cite{shoeybi2019megatron}}. 
%rely on Transformer structure in their optimization.
%, namely convolutional neural networks. 
%However, most of the memory demanding models come from machine translation and language modeling rather than computer vision.  

%Out-of-core methods: 

{\bf Out-of-core and Recomputation:} these methods either augment the device by utilizing an extra memory (Out-of-core methods) or optimize the memory consumption of the model (recomputation methods). vDNN~\cite{rhu2016vdnn} is a memory manager that virtualizes GPU memory in DNN training. ooc{\_}cuDNN~\cite{DBLP:conf/bigdataconf/ItoME17} extends cuDNN and applies cuDNN-compatible operators even when a layer exceeds GPU memory capacity by swapping at the granularity of individual tensor dimensions. Gradient checkpointing~\cite{Chen2016TrainingDN} %trades computation for memory, it 
reduces the memory needed to store the intermediate outputs and gradients with the cost of doubling the forward pass computational cost %This trade-off incurs up to roughly 30\% of additional runtime cost
~\cite{mlsys2020_196,Chen2016TrainingDN}. PoocH~\cite{Ito2019uPurofilingB} and Capuchin~\cite{10.1145/3373376.3378505} propose a hybrid approach that selects either recomputing or swapping for certain layers to reduce the performance overhead based on profiling data.

{\bf Graph partitioning}: 
%Even though not directly related to DL training, we present the relevant related work on graph algorithms.  
%Most existing  graph partitioning libraries are designed to handle undirected graphs. 
To deal with a directed graph, existing  graph partitioning libraries convert every directed edge to undirected even though this conversion  loses crucial information~\cite{bader2013graph}. 
%An extension to the general graph partitioning problem is to statically map a set of processes to processors so that the communication time  is minimized while balancing workloads~\cite{bichot2011graph}. A crucial distinction between the static mapping and the scheduling is that the scheduling considers the logical and temporal dependencies of processes, while the mapping assumes that all the processes simultaneously coexist during the entire execution.
Due to this reason, Scotch static mapper ~\cite{pellegrini1996scotch, pellegrini2009distillating} and MinCut optimizer, results in 2 to 10 times slowdown when applied on graphs of DL models~\cite{10.5555/3305890.3305932, mirhoseini2018hierarchical}. In~\cite{herrmann2017acyclic}, new techniques are proposed to deal with directed graphs and ~\cite{ozkaya2019scalable} built on top of those techniques for a clustering based scheduler. They aim at producing {\em acyclic partitioning}, where if there is a cut edge from partition $a$ to $b$ and another from $b$ to $a$, the partition is considered cyclic, and is not acceptable. Since the graphs produced by Tensorflow are full of fork-joins, applying their technique to our DNN models results in  unbalanced partitions.

{\bf Static graph scheduling:} Plenty of sophisticated and high-quality algorithms were proposed~\cite{kwok1995bubble, kwok1996dynamic, yang1994dsc, hwang1989scheduling, he2018novel} in this area. The vast majority of these algorithms were developed in 1990's to handle small-sized graphs, and they  were later evaluated using instances having up to 3000 nodes 
%tens or hundreds of nodes; less frequently exceeding 1000 and at maximum 3000 nodes
~\cite{wang2016comparative, liou1997comparison, he2018novel,wang2018list, gerasoulis1992comparison}. A recent evaluation on large graphs shows that they either do not scale due to their high time-complexity, or produce low-quality allocations due to their inability to capture the global structure of the graph~\cite{ozkaya2019scalable}. \edit{Table~\ref{tab:complexity_sched} shows the time complexity of some of these algorithms. Note that these heuristics are only scheduling heuristics (do not include memory handling component), hence their complexities should be compared against Step 1 of \algo.}

\begin{table}[t]
\begin{center}
\footnotesize{
\caption{\edit{Time complexity of some of the best task scheduling algorithms}}
%\wahib{I am assuming we add up the steps since they are sequential, I can understand that most will add up to the overall, except the mapping. Is there a proof that $K|V|+|E|$ is always lger than the mapping complex? and even then, shouldn't we use complexity bounds? or do you have something else in mind?} \wahib{Update: I changed it to ${O(\vert V \vert * log\vert V \vert) + K(\vert E \vert)}$, until we discuss again and decide otherwise} 
\label{tab:complexity_sched}

\begin{tabularx}{0.8\linewidth}{|l|X|}
\hline
Algorithm & Time Complexity \\\hline
\quad Dynamic Critical Path~\cite{kwok1996dynamic} & ${O(\vert V \vert^3)}$ \\
\hline
\quad Bubble Scheduling~\cite{kwok1995bubble} & ${O(K^2\vert V \vert \vert E \vert)}$ 
\\
\hline
\quad Earliest Task First~\cite{hwang1989scheduling} & ${O(K\vert V \vert^2)}$ \\
\hline
\quad Linear Clustering~\cite{kim1988general} & ${O(\vert V \vert(\vert V \vert + \vert E \vert))}$ \\
\hline
\end{tabularx}
}
\end{center}
\end{table}

%\didem{here briefly state how our approach differs from all the others discussed above. Why is it good? Why our algorithms do not suffer from these limitations}

\section{Conclusion}
\algo{} presents a lightweight and automatic approach to partition computational graphs of very large DNN models. It permits the training of models that do not fit into a single device memory, and enhances the training throughput of models barely fitting into a single device memory while being non-intrusive and generic. \algo{} is applied ahead of time and, hence the partitioning is available at the beginning of a run what enables applying any type of dynamic(runtime) optimizations on top of it. Due to the limited degree of parallelism of DNN graphs and the trend towards including more accelerators in one node, we proposed to apply \algo{} within a single node. Nevertheless, it can be used as an integral part in a large-scale training where Data Parallelism is used as an inter-node technique. The experiments on five large DNNs and comparisons with related work demonstrate its high efficiency and superlinear scaling of batch size and training throughput. 

\section*{Acknowledgement}
Authors from Koç University are supported by the Turkish Science and Technology Research Centre Grant No: 118E801. This work was partially supported by JST-CREST under Grant Number JPMJCR19F5. The research presented in this paper has benefited from the Experimental Infrastructure for Exploration of Exascale Computing (eX3), which is financially supported by the Research Council of Norway under contract 270053. Dr. Didem Unat is supported by the Young Scientist Awards Program by the Turkish Science Academy.

%% Acknowledgments
\begin{comment}
\begin{acks}                            %% acks environment is optional
                                        %% contents suppressed with 'anonymous'
  %% Commands \grantsponsor{<sponsorID>}{<name>}{<url>} and
  %% \grantnum[<url>]{<sponsorID>}{<number>} should be used to
  %% acknowledge financial support and will be used by metadata
  %% extraction tools.
  This material is based upon work supported by the
  \grantsponsor{GS100000001}{National Science
    Foundation}{http://dx.doi.org/10.13039/100000001} under Grant
  No.~\grantnum{GS100000001}{nnnnnnn} and Grant
  No.~\grantnum{GS100000001}{mmmmmmm}.  Any opinions, findings, and
  conclusions or recommendations expressed in this material are those
  of the author and do not necessarily reflect the views of the
  National Science Foundation.
\end{acks}
\end{comment}

%% Bibliography
%\bibliography{bibfile}

\bibliographystyle{elsarticle-num}
\bibliography{references}

%% Appendix
%%\appendix
%\newpage
%\section{Appendix}
%\input{appendix}
\end{document}